\documentclass[preprint,12pt]{elsarticle}

\usepackage{amsmath,amssymb,bm}
\usepackage{booktabs}
\usepackage{enumerate}
\usepackage{float}
\usepackage{subcaption}
\usepackage{tikz}
\usetikzlibrary{positioning,arrows.meta,fit,backgrounds,shapes.geometric}

\usepackage{xcolor}
\definecolor{ASTRAblue}{RGB}{31,90,158}
\definecolor{ASTRAnavy}{RGB}{20,42,84}
\definecolor{ASTRAorange}{RGB}{224,122,31}
\definecolor{ASTRAbglight}{RGB}{233,240,250}

\journal{Wave Motion}

\graphicspath{{./figs/}}

\newcommand{\mbf}[1]{\mathbf{#1}}
\newcommand{\be}{\begin{equation}}
\newcommand{\ee}{\end{equation}}
\newcommand{\mcl}[1]{\mathcal{#1}}

\begin{document}

\begin{frontmatter}
\title{Neural-field design of broadband Rayleigh-wave carpet cloaks
under microstructure realisability constraints}

\author[1,2]{David Aznaurov}
\author[1]{Davit Piliposyan\corref{cor1}}
\cortext[cor1]{Corresponding author}
\ead{piliposyan@mechins.sci.am}
\author[1]{Danila Rukhovich}
\author[3]{Sébastien Guenneau}

\affiliation[1]{organization={M3L Lab, Institute of Mechanics, 24B Baghramyan Ave., 0019 Yerevan},
                country={Armenia}}
\affiliation[2] {organization={Biutopic, 18 Rue Paul Ramadier, 44201 Nantes}, country={France}}
\affiliation[3]{organization={The Blackett Laboratory, Department of Physics \& UMI 2004 Abraham de Moivre-CNRS, Imperial College London}, addressline={London SW7 2AZ}, country={United Kingdom}}

\begin{abstract}

Transformation elasticity provides appropriate material distributions for elastodynamic cloaks but the required stiffness tensors generally violate the minor symmetries of  Cauchy elasticity and are difficult to realise using conventional materials.
Existing approaches restore these symmetries by modifying the transformed tensor, producing only an approximate cloak. 
In this work rather than modifying the transformed tensor we seek the best-performing cloak within the class of Cauchy materials. We formulate 2D Rayleigh wave carpet cloak design as an optimisation problem governed by partial differential equations. Using a coordinate based neural-field and a differentiable finite element model solver we optimise symmetric stiffness and density fields by minimising wave field distortion. Both single frequency and broadband optimisation are considered, with the broadband model trained over multiple frequencies. Physical realisability is addressed using a database of homogenised microstructures through conditional diffusion, neural-field inverse design, and nearest-neighbour selection. FEM simulations show that the optimised Cauchy design approaches the ideal transformation-based cloak. After projection onto explicit  microstructures, the homogenised representation recovers approximately $97\%$ of the defect-free reference surface-displacement magnitude, while direct FEM simulation of the fully resolved microstructured geometry recovers approximately $76\%$.
\end{abstract}

\begin{keyword}
elastic cloaking \sep Rayleigh waves \sep transformation elasticity \sep
neural field \sep implicit neural representation \sep metamaterials
\end{keyword}

\end{frontmatter}

\section{Introduction}\label{sec:intro}

Transformation elasticity provides a framework for steering elastic waves around an object by deriving the required material properties from a coordinate transformation. Specifically, a coordinate map $\mbf{x}=\Xi(\mbf{X})$ transforms the reference domain to create a cloaking region surrounding the object. The transformed elastodynamic equations then determine the spatially varying material properties required within this region to guide waves around the object while minimising disturbance to the exterior ~\citep{milton2006,brun2009,norris2011,diatta2014controlling}. It is the mechanical counterpart of
transformation optics~\citep{pendry2006}. The ``carpet cloak'' variant, which
hides a defect on an otherwise flat free surface~\citep{li2008} is the
configuration most accessible to experiments in the three-dimensional elastostatic setting ~\citep{buckmann2014elasto}. 
A two-dimensional mechanical cloak based on a direct lattice transformation was experimentally demonstrated in the static regime~\citep{buckmann2015}; however, numerical simulations indicate that its cloaking performance deteriorates markedly beyond the quasi-static regime~\citep{kadic2020elastodynamic}.

In \citep{brun2009,diatta2014controlling} the coating material follows from the Brun--Guenneau--Movchan (BGM)
push-forward transformation 
$c^{\mathrm{eff}}_{ijkl}=J^{-1}\,C_{IjKl}\,F_{iI}\,F_{kK}$ and
$\rho^{\mathrm{eff}}=\rho\,J^{-1}$, where $\mbf{F}=\nabla_{\!X}\mbf{x}$
is the transformation gradient and $J=\det\mbf{F}$~\citep{brun2009,norris2011}.
The resulting tensor keeps the major symmetry of classical elasticity but
breaks the minor symmetries, so the ideal cloak is a polar (Cosserat)
medium that lies outside the class of ordinary Cauchy
materials~\citep{milton1995,milton2006}. Turning this exact but
non-realisable prescription into a cloak that can be built from common
materials is the problem we address. Alternatively, one can enforce the symmetry of the transformed elasticity tensor, but this leads to an elasticity equation with an anisotropic mass density and additional rank-3 tensors, hard to achieve in practice~\citep{milton2006,norris2011}.

The constraints on the material parameters can be  relaxed substantially for two-dimensional elastic carpet cloaks, provided that the defect on the free surface is smooth~\citep{quadrelli2021elastic}. 
Motivated by this result, we focus on carpet cloaks for Rayleigh waves. 

The exact cloak is described by the polar tensor introduced above, while approaches to realising it using ordinary materials can be broadly divided into three families. The first is to symmetrise the tensor to obtain a genuine Cauchy material, at the cost of reducing cloaking to an approximate effect~\citep{norris2008,norris2011}. The second is to reproduce the broken minor symmetry using a purpose-built lattice that combines hinge-like spring-and-mass contacts with distributed restoring torques, producing a macroscopically polar medium~\citep{nassar2018}. The third is to select orthotropic Cauchy cells from a precomputed database, as done for quasi-static mechanical cloaks~\citep{wang2022}. For Rayleigh waves specifically,
\citet{chatzopoulos2023} compared triangular and semi-circular carpet
cloaks and symmetrised their tensors by the arithmetic mean, reporting the
symmetrised quadratic semi-circular design as the best performing.
The triangular cloak is especially attractive because its piecewise-affine transformation has a constant Jacobian within each half of the cloak. Consequently, the required effective material properties are homogeneous within each region.

Our approach to finding a design inside the Cauchy class of ordinary
materials is a differentiable pipeline built on a JAX-FEM solver~\citep{xu2020jaxfem}
that evaluates a wave-field cloaking objective, with gradients flowing
end-to-end through the FEM adjoint. We optimise
at a single frequency and across a target band. This
improves substantially on the symmetrised designs of
\citet{chatzopoulos2023}. The neural field acts as an implicit material
representation, of the kind used for scene reconstruction in computer
vision~\citep{mildenhall2020nerf,sitzmann2020siren,tancik2020fourier}.
Reparameterising a design field by a neural network and optimising it
through a differentiable solver has been used extensively in structural
topology optimisation, with convolutional~\citep{hoyer2019} and
coordinate-based, mesh-free networks~\citep{chandrasekhar2021,zehnder2021}.
Those methods minimise static compliance over a scalar density field; here
the network instead outputs an anisotropic stiffness tensor and density and
is trained against a frequency-domain elastodynamic cloaking objective, so
as to approximate the non-symmetric transformation-elasticity material.

We first use the pipeline to search for the best
single orthotropic Cauchy material that most closely reproduces the ideal
response, rather than fixing its stiffness parameters by symmetrisation. To improve on it while
staying realisable, we then discretise the cloak into a regular grid of
cells and let a coordinate-based neural field assign each cell its own
Cauchy moduli and density. We then take the design one step further, to an explicit realisation. Each
cell is filled with a two-phase (solid/void) microstructure drawn from a
database of homogenised unit cells~\citep{nakarmi2024}; the cells share a
single common base material and differ only in their internal geometry, so
the whole cloak can be manufactured from one constituent by varying the
pattern from cell to cell. 
To the best of our knowledge, no previous study has demonstrated a Rayleigh-wave elastodynamic cloak realised in this way from a common material: earlier work either stops at effective symmetrised
tensors~\citep{chatzopoulos2023}, or reproduces the exact polar tensor only
in principle with a special lattice, exact only for a circular cloak 
in the infinite-cell limit ~\citep{nassar2018}. 
We report both the gains of the microstructured cloak and the fraction 
of the ideal performance that survives homogenisation.

The paper is organised as follows. Section~\ref{sec:setup} sets up the
half-space problem, the BGM push-forward for the triangular carpet, the
cell discretisation, and the JAX-FEM solver and metrics.
Section~\ref{sec:method} introduces the neural-field design pipeline.
Section~\ref{sec:results} reports the single-frequency results against the
ideal and symmetrised baselines and extends the
design to a broadband objective. Section~\ref{sec:micro} constrains the
cloak to realisable microstructures and validates the single-material
realisation. Section~\ref{sec:discussion} discusses the results and their
limitations, and Section~\ref{sec:conclusion} concludes.

\section{Problem setup and FEM baseline}\label{sec:setup}

\begin{figure}[htbp]
  \centering
  \includegraphics[width=\linewidth]{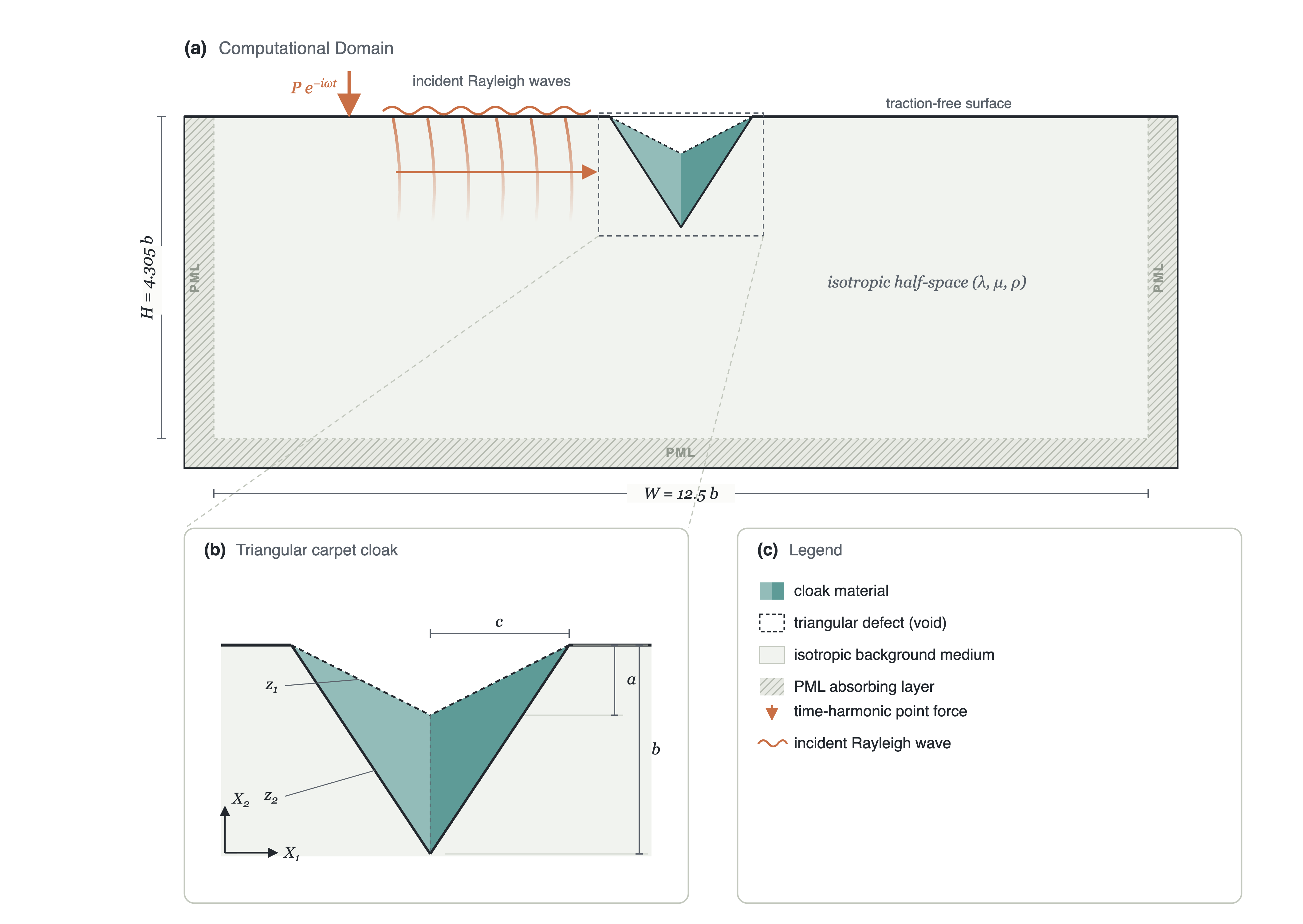}
  \caption{Problem configuration. \textbf{(a)}~Computational domain:
  an isotropic elastic half-space $(\lambda,\mu,\rho)$ of width
  $W=12.5\,b$ is truncated by Perfectly Matched Layers (PML) on the lateral and
  bottom boundaries; the top surface is traction-free. A time-harmonic
  vertical point force (red arrow) applied upstream generates Rayleigh
  waves (orange wavefronts) that propagate towards the triangular cloak
  (teal). \textbf{(b)}~Cloak geometry: the surface notch (void) has
  depth $a$ and half-width $c$; the outer cloak boundary sits at depth
  $b=3a$. Dashed lines mark the inner and outer triangle boundaries
  $z_1(X_1)$ and $z_2(X_1)$; the two mirror phases of the cloak are
  shaded in teal.}
  \label{fig:setup}
\end{figure}

\subsection{Governing equations}\label{subsec:governing}

We consider a homogeneous isotropic half-space with material properties
$(\lambda,\mu,\rho)$, where $\lambda$ and $\mu$ are the Lam\'e
coefficients and $\rho$ the mass density, and spatial coordinates
$\mbf{X}=(X_1,X_2)$ for the reference domain, with the free surface on
$X_2=0$. For in-plane surface waves, i.e.\ Rayleigh waves, propagating
along the horizontal $X_1$ direction, the governing Navier
elastodynamic equation takes the following form
\be\label{eq:navier_ref}
  \nabla_{\!X}\cdot\bigl(\mbf{C}:\nabla_{\!X}\mbf{U}\bigr)
  = \rho\,\mbf{U}_{tt},
\ee
where $\mbf{C}$ is the isotropic fourth-order elasticity tensor,
$\mbf{U}=(U_1,U_2)$ is the in-plane displacement, and $\mbf{U}_{tt}$ denotes its
second time derivative. Under the assumption of plane-strain
elasticity, the elastic tensor can be written in Voigt notation
$\{1,2,6\}=\{11,22,12\}$ as
\be\label{eq:Ciso}
  C_{IJ} =
  \begin{bmatrix}
    \lambda{+}2\mu & \lambda & 0 \\
    \lambda & \lambda{+}2\mu & 0 \\
    0 & 0 & \mu
  \end{bmatrix},
  \qquad I,J = 1,2,6.
\ee

We apply a pointwise invertible transformation $\Xi$ that maps the
reference configuration (virtual domain) $\mbf{X}\in\Psi$ to the
deformed region (physical domain) $\mbf{x}=\Xi(\mbf{X})\in\psi$ and the
remaining domain to itself. Given $\mbf{x}=(x_1,x_2)$ the coordinates
of the physical domain, the transformation gradient and its determinant
are
\be\label{eq:Fgrad}
  \mbf{F} = \nabla_{\!X}\,\mbf{x}
  =
  \begin{pmatrix}
    \dfrac{\partial x_1}{\partial X_1} & \dfrac{\partial x_1}{\partial X_2} \\[2mm]
    \dfrac{\partial x_2}{\partial X_1} & \dfrac{\partial x_2}{\partial X_2}
  \end{pmatrix},
  \qquad
  J = \det(\mbf{F}).
\ee
Equation~\eqref{eq:navier_ref} is not form-invariant under an arbitrary
transformation $\Xi$~\citep{milton2006}: the transformed material class
depends on the gauge $\mbf{U}(\Xi(\mbf{X}))=\mbf{A}\,\mbf{u}(\mbf{x})$
linking the virtual and physical displacements, with  a
non-singular matrix $\mbf{A}$. The choice $\mbf{A}=\mbf{F}$ leads to the Willis
setting, which guarantees a symmetric stress tensor but requires two
additional third-order coupling tensors and a tensorial mass
density~\citep{milton2006,norris2011,willis1981}, physically replicable
only within narrow frequency bands by resonant microstructures. For
this reason we adopt the Cosserat setting $\mbf{A}=\mbf{I}$, following
\citet{norris2011} and \citet{brun2009}, for which the governing
equation in the physical domain retains the Navier form
\be\label{eq:navier}
  \nabla_{\!x}\cdot\bigl(\mbf{c}^{\mathrm{eff}}:\nabla_{\!x}\mbf{u}\bigr)
  = \rho^{\mathrm{eff}}\,\mbf{u}_{tt},
\ee
with the transformed mechanical parameters given by the BGM transformation relations
\be\label{eq:pushforward}
  c^{\mathrm{eff}}_{ijkl} = J^{-1}\,C_{IjKl}\,F_{iI}\,F_{kK},
  \qquad
  \rho^{\mathrm{eff}} = \rho\,J^{-1}.
\ee
The transformed tensor preserves the major symmetries
($c^{\mathrm{eff}}_{ijkl}=c^{\mathrm{eff}}_{klij}$) but not the minor
ones,
\be\label{eq:broken_minor}
  c^{\mathrm{eff}}_{ijkl}
  \neq c^{\mathrm{eff}}_{jikl}
  \neq c^{\mathrm{eff}}_{ijlk}
  \neq c^{\mathrm{eff}}_{jilk},
\ee
except for special cases such as conformal transformations. The medium
is nonetheless described by a single fourth-order, non-symmetric inhomogeneous elastic tensor with scalar density: the
stress $\sigma_{ij}=c^{\mathrm{eff}}_{ijkl}\,\partial u_k/\partial x_l$
acts on the \emph{full} displacement gradient not the symmetric
strain and is itself non-symmetric (polar medium). All computations
below are carried out in the frequency domain with the time-harmonic
convention
$\mbf{u}(\mbf{x},t)=\operatorname{Re}\bigl[\mbf{u}(\mbf{x})\,e^{-\mathrm{i}\omega t}\bigr]$.

\subsection{Transformation elasticity for the triangular carpet cloak}
\label{subsec:triangular}

The cloak follows the pinched-carpet construction
of \citet{chatzopoulos2023}. With the free surface on $X_2=0$ and the
cloak centred at $X_1=0$, the inner (defect) and outer cloak boundaries
are the triangles
\be\label{eq:triangles}
  z_1(X_1) = \frac{a}{c}\,|X_1| - a,
  \qquad
  z_2(X_1) = \frac{b}{c}\,|X_1| - b,
\ee
where $a$ is the defect depth, $b$ the cloak depth, and $c$ the surface
half-width. The shear map
\be\label{eq:shearmap}
  x_1 = X_1,
  \qquad
  x_2 = \frac{z_2(X_1)-z_1(X_1)}{z_2(X_1)}\,X_2 + z_1(X_1)
\ee
compresses the triangular region above $z_2$ onto the cloak region
between $z_1$ and $z_2$, opening the notch. Its gradient is piecewise
constant,
\be\label{eq:Ftensor}
  \mbf{F} =
  \begin{bmatrix} 1 & 0 \\ F_{21} & F_{22} \end{bmatrix},
  \qquad
  F_{21} = \operatorname{sign}(X_1)\,\frac{a}{c},
  \qquad
  F_{22} = \frac{b-a}{b} = J.
\ee
For an isotropic background $(\lambda,\mu)$ the BGM
push-forward~\eqref{eq:pushforward} then yields, in augmented Voigt
ordering $[\sigma_{11},\sigma_{22},\sigma_{12},\sigma_{21}]$, the
effective stiffness
\be\label{eq:Ceff}
  \mbf{c}_{\mathrm{eff}}
  =
  \begin{bmatrix}
    \dfrac{2\mu{+}\lambda}{F_{22}} & \lambda & 0 &
      \dfrac{F_{21}}{F_{22}}(2\mu{+}\lambda) \\[2mm]
    \lambda & \dfrac{F_{21}^2\mu + F_{22}^2(2\mu{+}\lambda)}{F_{22}} &
      \dfrac{F_{21}}{F_{22}}\,\mu & F_{21}(\lambda{+}\mu) \\[2mm]
    0 & \dfrac{F_{21}}{F_{22}}\,\mu & \dfrac{\mu}{F_{22}} & \mu \\[2mm]
    \dfrac{F_{21}}{F_{22}}(2\mu{+}\lambda) & F_{21}(\lambda{+}\mu) &
      \mu & \dfrac{F_{21}^2(2\mu{+}\lambda) + F_{22}^2\mu}{F_{22}}
  \end{bmatrix},
\ee
together with the constant effective density
\be\label{eq:rhoeff}
  \rho^{\mathrm{eff}} = \frac{\rho}{J} = \frac{b}{b-a}\,\rho.
\ee
Unlike radial blow-up cloaks, whose tensors are singular at the inner
boundary, the triangular tensor is non-singular and uniform within each
half of the cloak (the shear $F_{21}$ changes sign across the
centreline $X_1=0$), so the ideal cloak consists of just two
mirror-image material phases. The price is that \emph{three} minor
symmetry pairs are broken independently
($c_{11,12}\neq c_{11,21}$, $c_{22,12}\neq c_{22,21}$,
$c_{12,12}\neq c_{21,21}$), and these are the entries that
symmetrisation strategies average away by hand.

\subsection{Realisable Cauchy parameterisation}\label{subsec:flat4}

Throughout this paper the designed cloak is restricted to orthotropic
Cauchy materials characterised by the four independent stiffness
parameters $(C_{11},C_{12},C_{22},C_{66})$, together with the scalar
mass density $\rho$, obtained by direct numerical search through the
differentiable FEM. This class is dictated by the microstructures that
must ultimately realise the cloak: square
two-phase unit cells invariant under the point group $D_2$, whose two
edge-aligned mirror axes fix the form of the homogenised stiffness.
Writing the plane-strain law in Voigt notation $\{1,2,6\}=\{11,22,12\}$,
a general symmetric Cauchy tensor has six independent constants,
\be\label{eq:cauchy_general}
  \begin{bmatrix}\sigma_{11}\\ \sigma_{22}\\ \sigma_{12}\end{bmatrix}
  =
  \begin{bmatrix}
    C_{11} & C_{12} & C_{16}\\
    C_{12} & C_{22} & C_{26}\\
    C_{16} & C_{26} & C_{66}
  \end{bmatrix}
  \begin{bmatrix}\varepsilon_{11}\\ \varepsilon_{22}\\ 2\varepsilon_{12}\end{bmatrix}.
\ee
Reflection across the edge, $(x_1,x_2)\mapsto(x_1,-x_2)$, leaves the normal strains unchanged
but reverses the sign of the shear strain. Mirror symmetry therefore eliminates
coupling between normal and shear deformation, requiring $C_{16}=C_{26}=0$.
The stiffness tensor consequently takes the block-diagonal orthotropic form
\be\label{eq:orthotropic}
  \mbf{C}_{\mathrm{ortho}}
  =
  \begin{bmatrix}
    C_{11} & C_{12} & 0\\
    C_{12} & C_{22} & 0\\
    0 & 0 & C_{66}
  \end{bmatrix}.
\ee
whose four moduli are exactly the per-cell design variables. We collect the four independent stiffness parameters into the vector
\be\label{eq:stiffness_vector}
  \mbf{p}_c
  =
  (C_{11},C_{12},C_{22},C_{66})^{\mathsf T}
  \in\mathbb{R}^{4}.
\ee
Thus, $\mbf{C}_{\mathrm{ortho}}$ denotes the orthotropic stiffness
matrix, while $\mbf{p}_c$ denotes its four independent stiffness
parameters. The mass density $\rho$ is treated separately.

\subsection{Cell discretisation}

\begin{figure}[htbp]
  \centering
  \includegraphics[width=0.75\linewidth]{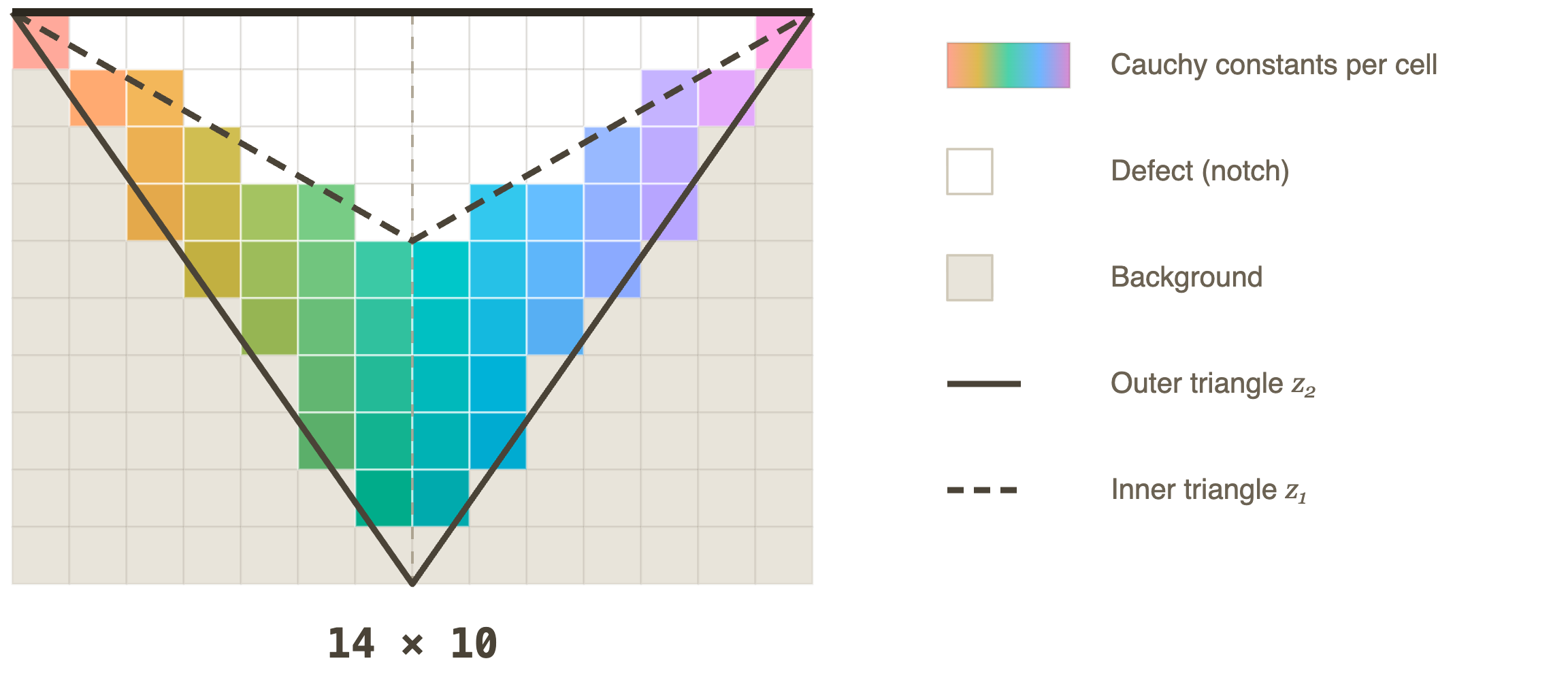}
  \caption{Cartesian cell decomposition of the triangular cloak at the
  $14\times10$ resolution used in all single-frequency experiments.
  Coloured cells carry per-cell stiffness parameters  
  $\mbf{p}_c$ and density $\rho$; white cells lie inside the
  defect (void notch); grey cells are background. The solid line is the
  outer cloak boundary $z_2$ and the dashed line is the inner
  boundary $z_1$.}
  \label{fig:cartesian_cells}
\end{figure}

We discretise the cloak region into a regular Cartesian grid of
$n_x\times n_y$ square cells (Fig.~\ref{fig:cartesian_cells}). Cells
whose centre lies between the inner and outer triangles
\eqref{eq:triangles} are assigned a piecewise-constant effective
stiffness and the density~\eqref{eq:rhoeff}. Because the triangular
transformation is affine in Cartesian coordinates, the
tensor~\eqref{eq:Ceff} is passed to the FEM solver without any rotation
to a local frame.

\subsection{Simulation domain and FEM solver}\label{subsec:domain}

We consider a rectangular computational domain of width $W=12.5\,b$ and
depth $H=4.305\,b$, truncating the half-space. The top boundary is
traction-free; Perfectly Matched Layers (reflectionless absorbing layers) terminate the two lateral
boundaries and the bottom. A triangular surface notch of depth
$a=0.0774\,H$ and half-width $c=0.1545\,H$ is centred on the free
surface, coated by a cloak of depth $b=3a$. The background medium is
soil, modelled as an isotropic elastic half-space with
$\rho=1600\,\text{kg}/\text{m}^3$, shear speed
$c_s=300\,\text{m/s}$, and $c_p=\sqrt{3}\,c_s$
($\nu=1/4$, $\lambda=\mu=144\,\text{MPa}$, plane strain). The
excitation is a time-harmonic vertical point force applied on the free
surface upstream of the cloak, generating Rayleigh waves that propagate
along the surface towards the defect. Frequencies are reported in the
dimensionless form $f^\star = f\,b/c_R$, where $c_R\approx0.92\,c_s$ is
the Rayleigh wave speed of the background: at $f^\star=1$ the Rayleigh
wavelength equals the cloak depth $b$.

The frequency-domain elastodynamic equations are solved with
complex-valued degrees of freedom and Rayleigh-damping PML through a
JAX-FEM finite-element solver~\citep{xu2020jaxfem} on a triangular mesh
with distance-based refinement near the cloak boundaries
($\sim\!246{,}000$ elements, $\sim\!124{,}000$ nodes). The entire
pipeline from cell material parameters through assembly and sparse
solve to the cloaking loss is differentiable via implicit adjoint
differentiation through the linear FEM solve, enabling gradient-based
optimisation of the per-cell stiffness and density.

\section{Neural-field design pipeline}\label{sec:method}

\subsection{Coordinate-conditioned material network}

\begin{figure}[htbp]
  \centering
  \begin{tikzpicture}[font=\footnotesize,>={Latex[length=2mm]},node distance=6mm,
      box/.style={draw=ASTRAblue,fill=ASTRAbglight,rounded corners=2pt,thick,
                  inner sep=4pt,align=center,minimum height=9mm}]
    \node[box] (xy)  {cell coords\\$(x,y)$};
    \node[box,right=of xy] (nf) {neural field\\$\Phi_\theta$};
    \node[box,right=of nf] (mat) {material\\$(\mbf{p}_c,\rho)$};
    \node[box,right=of mat] (fem) {JAX-FEM\\wave solve};
    \node[box,right=of fem] (loss) {cloak loss\\$\mcl{L}$};
    \draw[->,thick,ASTRAnavy] (xy)--(nf);
    \draw[->,thick,ASTRAnavy] (nf)--(mat);
    \draw[->,thick,ASTRAnavy] (mat)--(fem);
    \draw[->,thick,ASTRAnavy] (fem)--(loss);
    \draw[->,thick,dashed,ASTRAorange] (loss.south) to[out=-90,in=-90]
          node[below,midway]{adjoint gradient $\partial\mcl{L}/\partial\theta$} (nf.south);
  \end{tikzpicture}
  \caption{Neural-field design pipeline. Cell-centre coordinates are
  mapped through a Fourier positional encoding into a four-layer MLP
  that predicts the four stiffness parameters $\mbf{p}_c$
  and the density $\rho$. The piecewise-constant material assignment is expanded to
  the FEM quadrature points and the cloaking objective is differentiated
  back through the implicit JAX-FEM adjoint into the network weights.}
  \label{fig:nf-arch}
\end{figure}
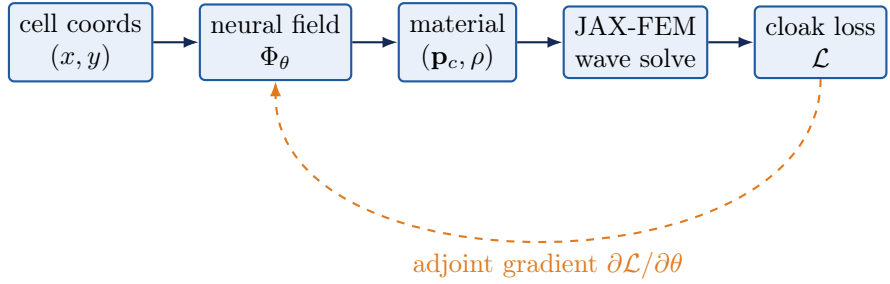

We reparameterise the cloak material field through a coordinate-based
multilayer perceptron (MLP)
\be\label{eq:neural_field}
  \Phi_\theta\colon(x,y)
  \;\mapsto\;
  \bigl(\mbf{p}_c(x,y),\,\rho(x,y)\bigr),
  \qquad
  \mbf{p}_c\in\mathbb{R}^{4}.
\ee
where $\theta$ denotes the network weights, $\mbf{p}_c$ is the
stiffness parameter vector defined in Eq.~\eqref{eq:stiffness_vector},
and $\rho>0$ is the scalar mass density. The network
is evaluated at each cloak-cell centre to produce the piecewise-constant
material assignment that is then expanded to the FEM quadrature points.
A four-layer MLP with 256 hidden units ($\sim\!200{,}000$ weights) is
used in all experiments; we adopt $\tanh$ activations and the Fourier
positional encoding of \citet{tancik2020fourier} so that the network
can represent sharp material transitions near the cloak boundaries.
Figure~\ref{fig:nf-arch} shows the architecture and the data flow
through the differentiable FEM. For optimisation initialisation, we considered two starting points: the arithmetic-mean symmetrisation of the ideal tensor~\eqref{eq:Ceff} and the mean of the microstructure dataset. We found no appreciable
difference in the converged results, indicating that the design is
driven primarily by the cloaking objective rather than by the
initialisation.

\subsection{Physical admissibility of the material output}\label{sec:constrained}

In our approach nothing by itself guarantees 
a realisable material; the MLP can drive
a stiffness negative, break the positive-definiteness of the Cauchy
tensor, or reach an anisotropy ratio no real microstructure can realise.
Rather than police these violations with penalty terms after the fact, we
build them into the decoding, so that every network output is a
physically admissible material by construction.

Per cloak cell, the network emits five unbounded raw channels
\[
\mbf{r}
=
(r_{11},r_{22},r_{66},r_{12},r_{\rho})
\in\mathbb{R}^{5}.
\]
The first four are mapped to the stiffness parameters through three
guarantees, while the fifth controls the density.

\emph{(i) Positive diagonal and shear.} The three positive-definite
entries are set by a log-space (hence strictly positive) multiplicative
residual,
\be\label{eq:constr_diag}
  C_{ii} = C_{ii}^{(0)}\,\exp(\epsilon\,r_{ii}),
  \qquad ii\in\{11,22,66\},
\ee
which keeps them positive for any real $r_{ii}$ and recovers
$C_{ii}^{(0)}$ at $\mbf{r}=\mbf 0$.

\emph{(ii) Bounded anisotropy.} To exclude ratios beyond what the
microstructure can attain, the diagonal ratio $C_{11}/C_{22}$ is confined
to $[1/R,\,R]$ while its geometric mean $\sqrt{C_{11}C_{22}}$ is held
fixed. Writing $g=\tfrac12\log(C_{11}/C_{22})$, a $\tanh$ squash on the
half-log ratio,
\be\label{eq:constr_aniso}
  g_b = \tfrac12\log R\,\tanh\!\Bigl(\tfrac{g}{\tfrac12\log R}\Bigr),
  \qquad
  C_{11},C_{22}=\sqrt{C_{11}C_{22}}\;e^{\pm g_b},
\ee
bounds the ratio smoothly to $[1/R,R]$ with $R$ the anisotropy cap (we
use $R=30$, chosen above the $99$th percentile of the microstructure
dataset with margin).

\emph{(iii) Positive-definite coupling.} The off-diagonal $C_{12}$ is
written through a correlation coefficient $\chi\in(-\kappa,\kappa)$
rather than directly,
\be\label{eq:constr_c12}
  C_{12}=\chi\,\sqrt{C_{11}C_{22}},
\qquad
\chi=\kappa\,\tanh\!\bigl(r_{12}+\phi_0\bigr),
\ee
with $\kappa<1$ (we use $\kappa=0.99$). Because $|\chi|<\kappa<1$, the
orthotropic block stays strictly positive-definite,
$\det=C_{11}C_{22}(1-\chi^2)>0$, for any $r_{12}$, and $C_{12}$ is free
to change sign. The offset $\phi_0=\operatorname{arctanh}(\chi^{(0)}/\kappa)$
centres the map so that $\mbf{r}=\mbf 0$ again recovers the initial
coupling $C_{12}^{(0)}$. Non-cloak cells bypass the decoding and keep
their background values exactly.

\emph{(iv) Positive density.} The density is decoded using the same
log-space multiplicative residual as the positive stiffness components,
\be\label{eq:constr_rho}
  \rho = \rho^{(0)}\,\exp(\epsilon\,r_{\rho}),
\ee
where $\rho^{(0)}>0$ is the initial density used for the corresponding
optimisation and $\epsilon$ is the same output scale as in
Eq.~\eqref{eq:constr_diag}. Since $\exp(\cdot)>0$, this parameterisation
guarantees $\rho>0$ for any real $r_{\rho}$ and recovers
$\rho^{(0)}$ when $r_{\rho}=0$. No explicit upper or lower magnitude
bound is imposed on $\rho$ by this decoding. In the
microstructure-constrained optimisation, compatibility with the
density range of the microstructure catalogue is additionally
regularised through the joint GMM prior on $(\mbf{p}_c,\rho)$ described
in Sec.~\ref{subsec:micro_pipeline}.

Together, \eqref{eq:constr_diag}--\eqref{eq:constr_rho} provide a smooth,
everywhere-differentiable decoding of the raw network outputs into a
positive-definite, bounded-anisotropy Cauchy stiffness and a strictly
positive density.

\subsection{Loss function}

The optimisation minimises a
\emph{magnitude-band integral}: the mean-squared relative error of the
displacement magnitude over a strip of depth $d=0.5\,b$ (half the cloak
depth) below the free surface, downstream of and excluding the
cloak/defect footprint,
\begin{subequations}\label{eq:opt_loss_cloak}
\begin{align}
  \mcl{L}_{\mathrm{cloak}}(\theta)
  &= \frac{1}{|\mcl{B}|}\int_{\mcl{B}}
      \left(\frac{|\mbf{u}_{\mathrm{cloak}}|}{|\mbf{u}_{\mathrm{ref}}|}-1\right)^{2}
      \,\mathrm{d}A, \label{eq:opt_loss_cloak_L}\\
  \mcl{B}&=\bigl\{(x,y)\in\Omega_{\mathrm{phys}}\setminus\overline{\Omega}_{\mathrm{cloak}}
     : y_{\mathrm{top}}-d\le y\le y_{\mathrm{top}},\ x>x_c\bigr\}.
     \label{eq:opt_loss_cloak_B}
\end{align}
\end{subequations}
where $\mbf{u}_{\mathrm{cloak}}$ and $\mbf{u}_{\mathrm{ref}}$ are the
cloaked and defect-free reference fields and $x_c$ is the downstream edge
of the cloak. Comparing magnitudes rather than the complex fields
themselves makes the objective phase-invariant, which is deliberate here: a wave
that is diverted around the defect follows a slightly longer path and
arrives with a small phase lag relative to the flat-surface reference.
An explicit
neighbour-smoothness term used in raw-parameter
baselines~\citep{wang2022} is unnecessary here: the neural field is
spatially continuous by construction, and its smoothness is controlled
directly through the number and bandwidth of the Fourier positional
features~\citep{tancik2020fourier}.

While~\eqref{eq:opt_loss_cloak} is what the optimiser minimises, performance
is \emph{reported} through the dimensionless \emph{cloak ratio}
\be\label{eq:cloak_ratio}
  \eta \;=\;
  \frac{\langle|\mbf{u}|\rangle}{\langle|\mbf{u}_{\mathrm{ref}}|\rangle}
\ee
of the mean surface-displacement magnitude on the free surface beyond the
cloak footprint to that of the defect-free reference, following
\citet{chatzopoulos2023}, so that $\eta\to1$ is perfect cloaking.

\section{Results}\label{sec:results}

Unless stated otherwise, results are reported at the design frequency
$f^\star=2.0$ through the cloak ratio~\eqref{eq:cloak_ratio}. We build up
the design in steps of increasing material freedom from a single
homogeneous realisable material to a full cell grid
(Secs.~\ref{subsec:four_materials}--\ref{subsec:final14}) then widen the
objective from one frequency to a band (Sec.~\ref{sec:multifreq}) and verify
that the resulting medium is physically admissible
(Sec.~\ref{subsec:bloch}).

\subsection{How many distinct materials does the cloak need?}\label{subsec:four_materials}

We first ask how far a \emph{single} realisable material can go. Both
cloak phases are collapsed onto one orthotropic Cauchy material, a
$1\times1$ cell filling the whole cloak, whose stiffness parameters
$\mbf{p}_c$ and density $\rho$ are optimised through the differentiable
FEM at $f^\star=2.0$. The natural point of comparison is the closed-form arithmetic-mean symmetrisation of \citet{chatzopoulos2023}, instantiated and solved on the same mesh. That baseline is in fact richer than the tile it is compared against: averaging each
broken pair to its mean (rather than annihilating it) keeps the normal--shear
coupling $(C_{16},C_{26})$, so the symmetrised medium is a fully
anisotropic Cauchy material, and, since $F_{21}$ only changes sign across
the centreline, it is \emph{two} such mirror phases. The single orthotropic material therefore competes with strictly less freedom: one orthotropic tensor against two
anisotropic ones.

Table~\ref{tab:single_vs_sym} reports the outcome. The symmetrised tensor
recovers only $\eta=0.63$ and distorts the near field more than the bare notch,
whereas a single optimised orthotropic tile reaches $\eta=0.86$, a
substantial move towards the ideal continuous polar cloak
($\eta=0.999$) using a smaller material class and half as many
materials. This provides clear evidence that optimising inside the
Cauchy class outperforms symmetrisation.
The advantage is concentrated around the design frequency, as expected for a
single-frequency optimisation. Matching the baseline's fully anisotropic Cauchy class and material count with two optimised materials, one per half, makes the comparison
like-for-like and lifts the cloak ratio to $\eta=0.947$.

The fully anisotropic designs are included only for comparison with the
symmetrisation baseline. They are not realisable by the
$D_2$-symmetric square cells of Sec.~\ref{sec:micro}, so the remainder
of the paper focuses on the orthotropic Cauchy class that the
microstructure catalogue can realise.

\begin{table}[htbp]
\centering
\caption{Single-frequency cloak ratio $\eta$ at $f^\star=2.0$ for
homogeneous Cauchy materials filling the cloak. The orthotropic class
has the stiffness form~\eqref{eq:orthotropic}, while the fully
anisotropic Cauchy class additionally admits the normal-to-shear
coupling $(C_{16},C_{26})$. A single optimised orthotropic material
outperforms the symmetrisation of \citet{chatzopoulos2023} from a
smaller material class and using half as many materials. Matching the
symmetrisation baseline's material class and material count with two
fully anisotropic materials reaches $\eta=0.947$. All optimised rows
share the mesh, source, loss, and design frequency. The fully
anisotropic materials are not realisable with the $D_2$ square cells
of Sec.~\ref{sec:micro}.}
\label{tab:single_vs_sym}
\begin{tabular*}{\textwidth}{@{\extracolsep{\fill}}lccc@{}}
\toprule
Configuration & Materials & Class & $\eta$ \\
\midrule
Symmetrised ideal (Chatzopoulos)       & 2 & Anisotropic Cauchy & $0.634$ \\
Single optimised material ($1\times1$) & 1 & Orthotropic Cauchy & $0.857$ \\
Optimised left/right pair ($2\times1$) & 2 & Anisotropic Cauchy & $\mathbf{0.947}$ \\
\midrule
Ideal continuous $\mbf{c}_{\mathrm{eff}}$ & 2 & Polar & $0.999$ \\

\bottomrule
\end{tabular*}
\end{table}

Giving each phase more freedom closes most of the remaining gap. Splitting
the cloak into a coarse $n_x\times n_y$ grid of independent orthotropic tiles
and optimising all of them jointly, we find that even a $2\times2$
decomposition, using just \emph{four} realisable materials, lifts the cloak
ratio to $\eta=0.962$, i.e.\ within $4\%$ of a perfect cloak. 

\subsection{Grid selection}
\label{subsec:grid_selection}

Refining the grid further, we selected the resolution by weighing cloaking
performance against optimisation cost. Materials are assigned only to cells
whose centre falls between the inner and outer triangles~\eqref{eq:triangles},
so an $n_x\times n_y$ grid uses far fewer than $n_x n_y$ materials. The
$14\times10$ grid nominally spans $140$ cells but places only $46$ inside the
cloak region; these $46$ are the actual design variables. It proved optimal in
this trade-off: cloaking is already converged at $\eta\approx0.99$, while
finer grids only multiply the solve cost without improving on it. We adopt
$14\times10$ for all subsequent single-frequency work.

\subsection{Near-perfect single-frequency cloaking}\label{subsec:final14}

\begin{figure}[htbp]
  \centering
  \includegraphics[width=0.82\linewidth]{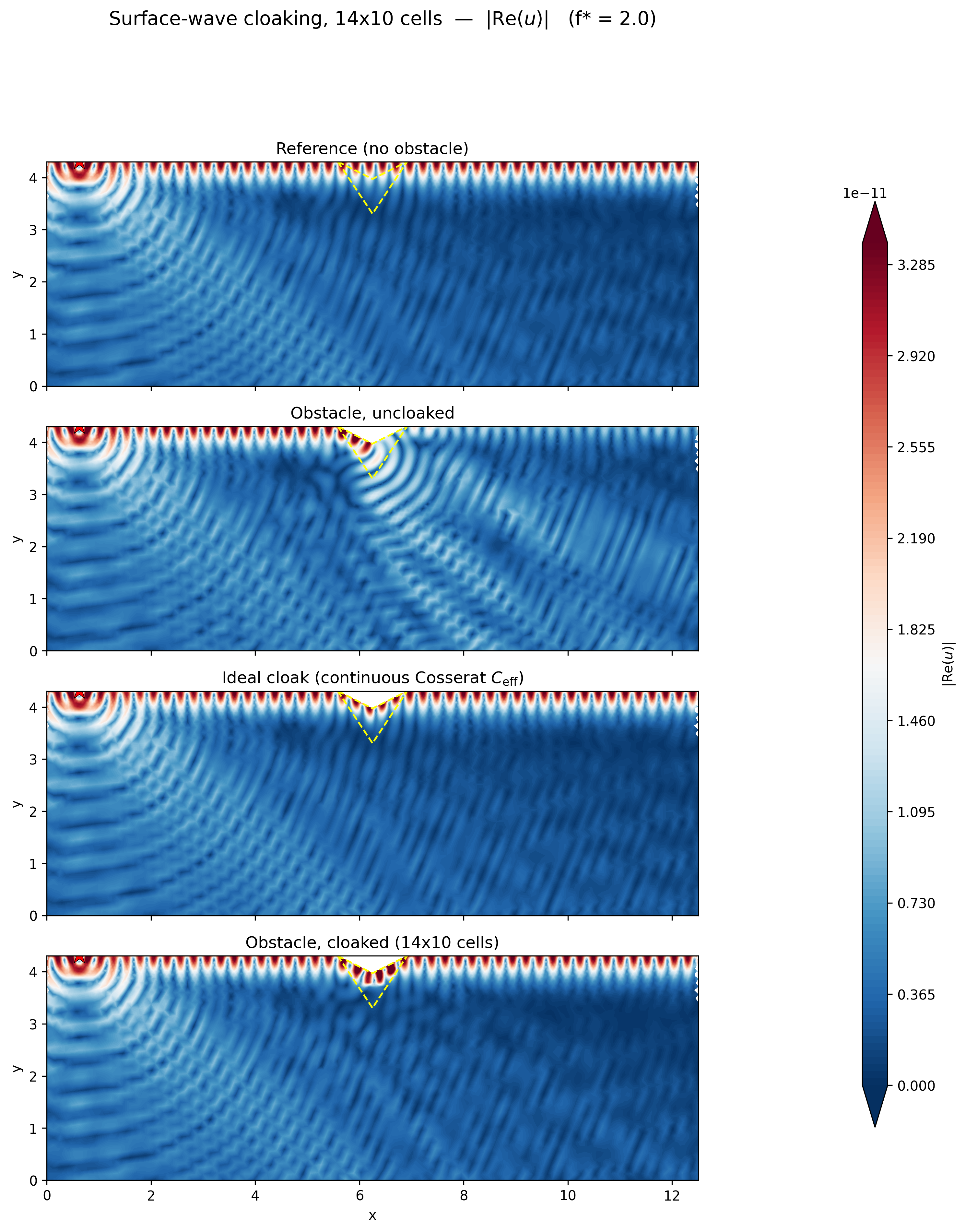}
  \caption{Real-displacement magnitude $|\!\operatorname{Re}\mbf{u}|$ at
  $f^\star=2.0$ (top to bottom): defect-free reference, uncloaked notch,
  ideal continuous polar cloak $\mbf{c}_{\mathrm{eff}}$, and the optimised
  $14\times10$-cell cloak ($46$ in-cloak materials, $\eta=0.99$). The optimised cloak closely reproduces the reference displacement-amplitude
pattern on the transmission side, with the downstream shadow of the uncloaked notch removed. Dashed lines
  mark the cloak/defect boundary.}
  \label{fig:single14_field}
\end{figure}

\begin{figure}[htbp]
  \centering
  \includegraphics[width=\linewidth]{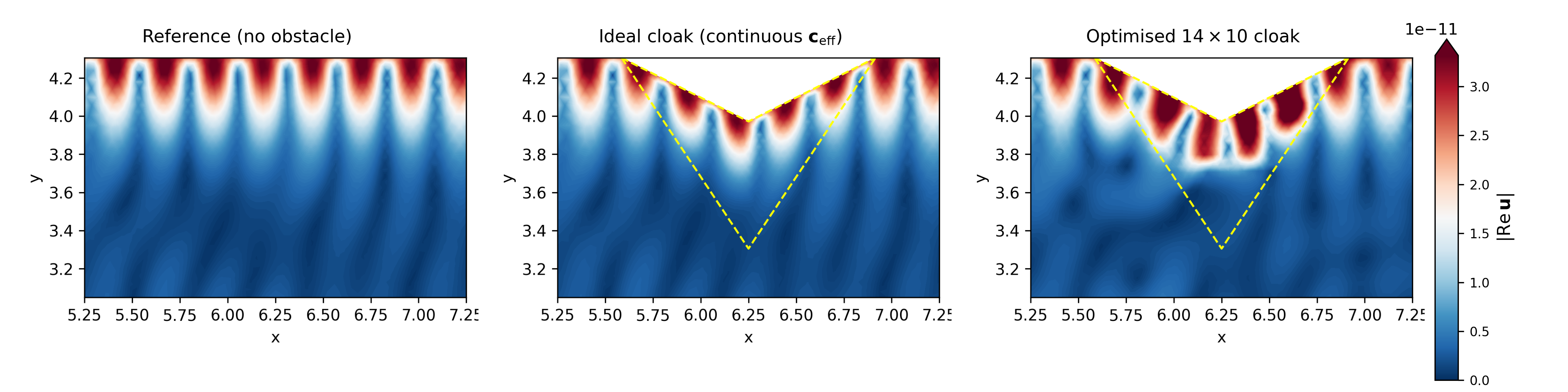}
  \caption{Zoom on the cloak region ($|\!\operatorname{Re}\mbf{u}|$ at
  $f^\star=2.0$, shared colour scale): defect-free reference (left), ideal
  continuous polar cloak (centre), and optimised $14\times10$ cloak (right);
  dashed lines outline the inner (defect) and outer cloak boundaries. All three
  reproduce the same amplitude envelope, but the optimised cloak leaves its
  near-surface wavefronts slightly phase-shifted ($\approx0.18\lambda$,
  $\approx65^\circ$) relative to the reference and the ideal cloak, corresponding to a
constant phase offset that is not penalised by the phase-invariant
magnitude objective.}
  \label{fig:single14_zoom}
\end{figure}

\begin{figure}[htbp]
  \centering
  \includegraphics[width=\linewidth]{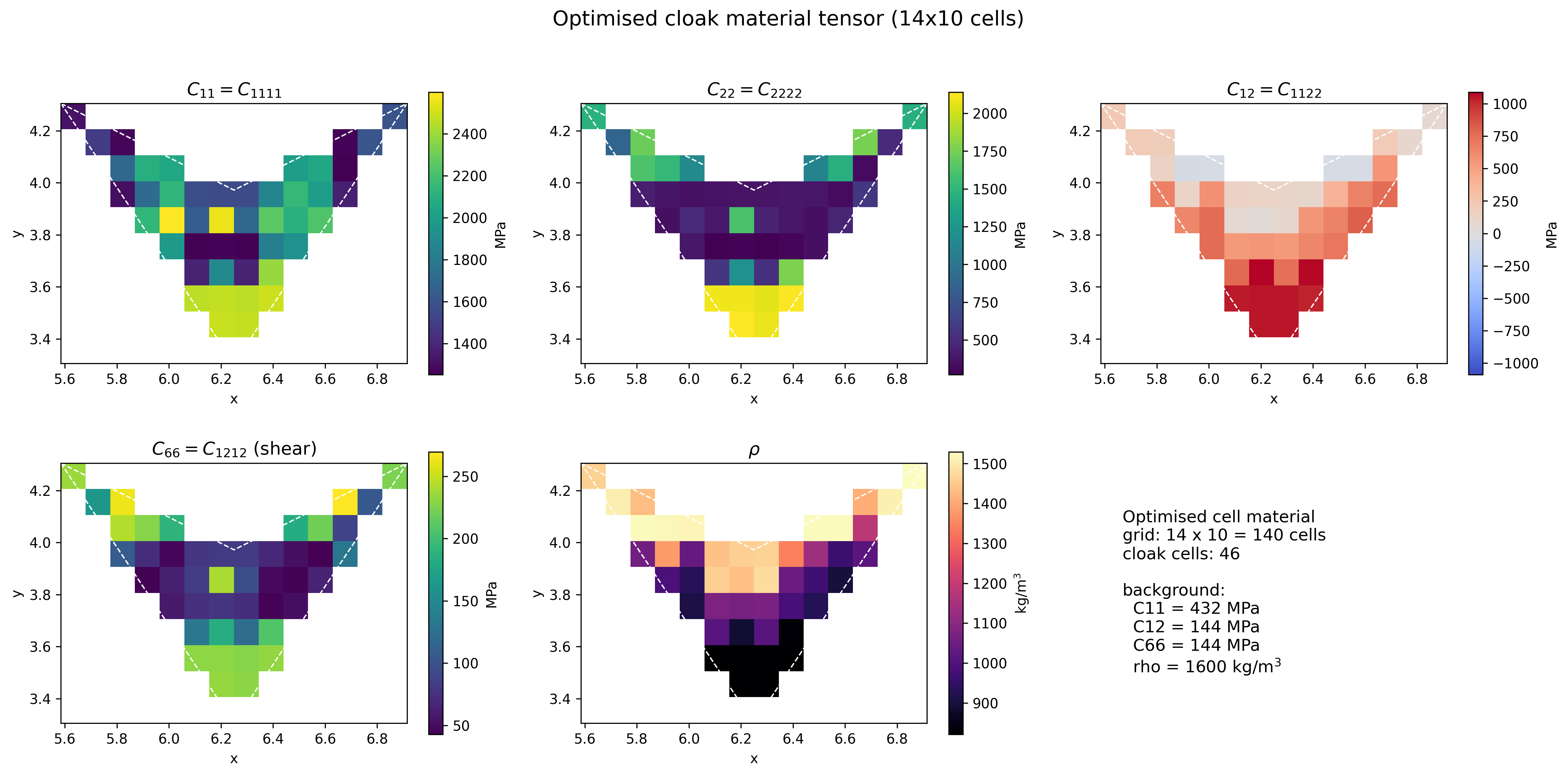}
  \caption{Optimised per-cell material of the $14\times10$ single-frequency
  cloak: the four orthotropic stiffness parameters $\mbf{p}_c$
and the density $\rho$ over the $46$ cells inside the cloak region (background
  cells masked). Only these $46$ materials are optimised.}
  \label{fig:single14_material}
\end{figure}

The adopted $14\times10$ design ($46$ optimised in-cloak orthotropic Cauchy materials) reaches $\eta=0.990$, effectively closing the continuum-level performance gap to the ideal continuous cloak.
Figure~\ref{fig:single14_field} compares the real-displacement magnitude of
the optimised cloak against three references on the same mesh: the defect-free
reference, the uncloaked notch, and the ideal continuous polar
$\mbf{c}_{\mathrm{eff}}$. The uncloaked notch back-scatters the incident
Rayleigh wave and casts a clear shadow downstream; the optimised $14\times10$
cloak restores the surface wavefield to the reference, matching the ideal
continuous cloak panel while using only realisable orthotropic Cauchy
materials. The only visible residual is a slight phase shift of the optimised solution:
Figure~\ref{fig:single14_zoom} zooms into the cloak region and shows that,
whereas the ideal continuous cloak keeps its near-surface wavefronts
phase-aligned with the reference, the optimised $14\times10$ cloak carries them
$\approx0.18\lambda$ ($\approx65^\circ$) later.  This phase lag is consistent with the phase-invariant magnitude
loss~\eqref{eq:opt_loss_cloak}, which constrains the displacement
magnitude but does not penalise phase differences. The resulting
constant phase offset, associated with the additional path length of a
wave routed around the notch, is therefore left unpenalised. Figure~\ref{fig:single14_material} shows the corresponding
optimised per-cell stiffness parameters $\mbf{p}_c$ and density $\rho$
over the $46$ cloak cells.

\subsection{Broadband multifrequency optimisation}\label{sec:multifreq}

The designs above are optimised at
one frequency and, being tuned to it, cloak well only in a narrow neighbourhood
of $f^\star=2$: away from the design frequency the cloak ratio falls off
markedly. Because the ideal triangular
tensor~\eqref{eq:Ceff} is frequency-independent and cloaks exactly at every
frequency, this narrowness is a limitation of the single-frequency
\emph{objective}, not of the material class. We therefore train a single design
against a band of frequencies at once.

The broadband loss is the average of the single-frequency magnitude-band
integral~\eqref{eq:opt_loss_cloak} over a finite set $\mcl{S}$ of frequencies
sampling the target band,
\be\label{eq:band_loss}
  \mcl{L}_{\mathrm{band}}(\theta)
  = \frac{1}{\textstyle\sum_{f\in\mcl{S}}w_f}
    \sum_{f\in\mcl{S}} w_f\,\mcl{L}_{\mathrm{cloak}}(\theta;f),
\ee
where $\mcl{L}_{\mathrm{cloak}}(\theta;f)$ is~\eqref{eq:opt_loss_cloak}
evaluated at frequency $f$ against the corresponding defect-free reference field
$\mbf{u}_{\mathrm{ref}}(f)$. Equal weights $w_f=1$ minimise the mean in-band
distortion; the weights are free to emphasise particular frequencies, and a
worst-case (min--max) variant that up-weights the currently least-cloaked
frequency can be used to flatten the response across the band. A single neural
field $\Phi_\theta$ produces the shared per-cell material, so the same design is
required to cloak at every $f\in\mcl{S}$ simultaneously.

The objective~\eqref{eq:band_loss} places no restriction on which band is
targeted; the only cost is computational, since every optimisation step
evaluates a forward and an adjoint FEM solve per frequency and the per-step
cost therefore scales linearly with $|\mcl{S}|$.

\begin{figure}[htbp]
  \centering
  \includegraphics[width=0.9\linewidth]{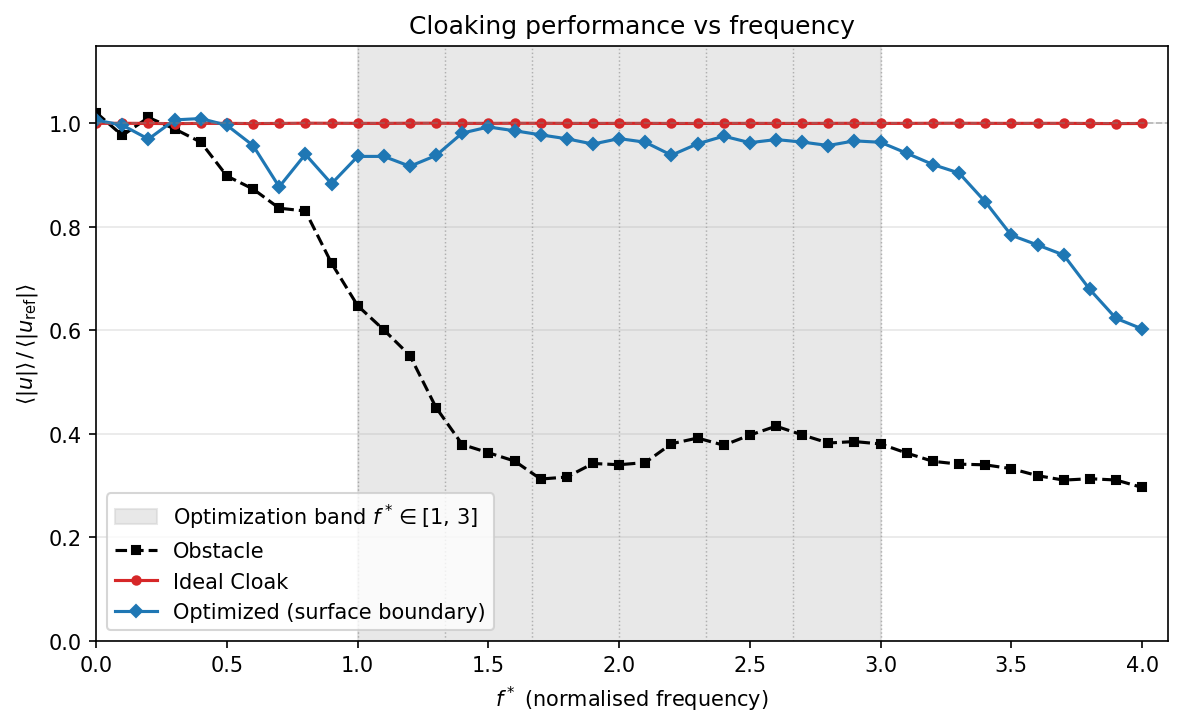}
  \caption{Frequency sweep of the cloak ratio
  $\eta=\langle|\mbf{u}|\rangle/\langle|\mbf{u}_{\mathrm{ref}}|\rangle$ for the
  broadband $14\times10$ design trained on the band $f^\star\in[1,3]$ (shaded):
  the ideal continuous transformation cloak (red, $\eta\approx1$ at every
  frequency), the uncloaked notch (black, dashed), and the broadband optimised
  cloak (blue). The optimised design holds $\eta\approx0.96$ across the entire
  training band and degrades gracefully outside it.}
  \label{fig:multifreq_sweep}
\end{figure}

We optimise the same $14\times10$ design ($46$ in-cloak Cauchy materials)
against the band $f^\star\in[1,3]$, sampled at seven uniformly spaced
frequencies with equal weights, and evaluate the converged design on a dense
sweep $f^\star\in[0,4]$ (Fig.~\ref{fig:multifreq_sweep}). The single design
holds $\eta=0.92$--$0.99$ (mean $0.96$) across the whole training band, where
the uncloaked notch falls to $\eta\approx0.31$, and degrades gracefully outside
it ($\eta\approx0.78$ at $f^\star=3.5$, $\approx0.60$ at $f^\star=4.0$). One
realisable design therefore retains most of the single-frequency performance
over an octave-plus band rather than at one frequency only.

\subsection{Floquet--Bloch dispersion}\label{subsec:bloch}

\begin{figure}[htbp]
  \centering
  \begin{subfigure}{0.49\textwidth}
    \centering
    \includegraphics[width=\linewidth]{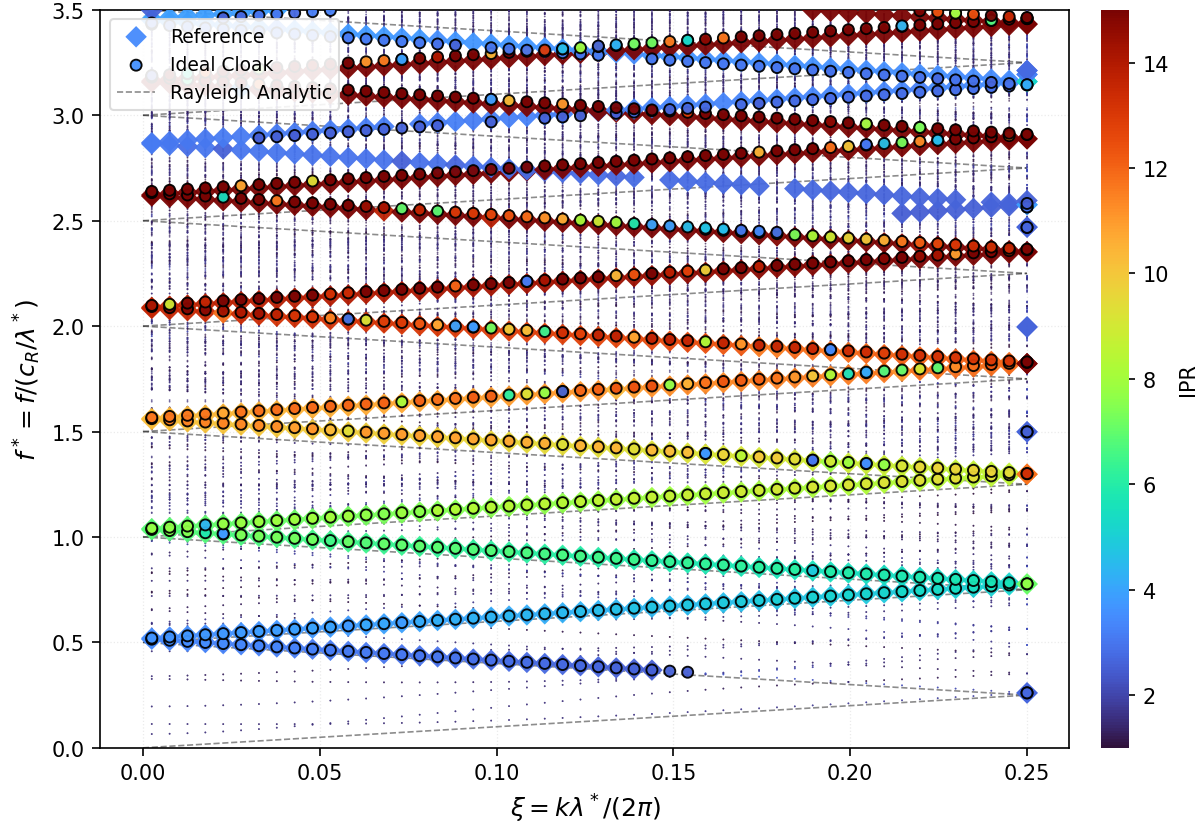}
    \caption{Analytical transformation cloak}
    \label{fig:bloch_analytical}
  \end{subfigure}\hfill
  \begin{subfigure}{0.49\textwidth}
    \centering
    \includegraphics[width=\linewidth]{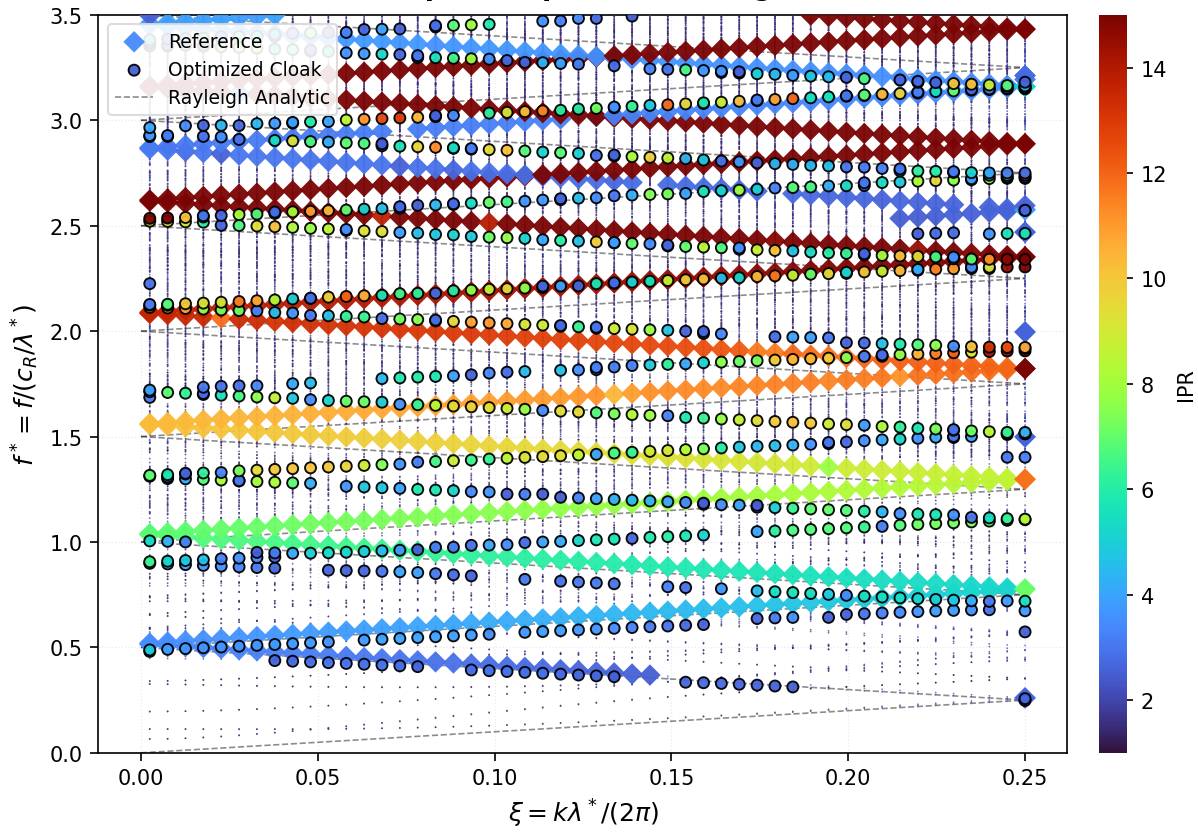}
    \caption{Optimised ($14\times10$) neural-field cloak}
    \label{fig:bloch_optimized}
  \end{subfigure}
  \caption{Floquet--Bloch dispersion diagrams for the full-cloak unit
  cell described above. In each panel, the uncloaked reference cell
  (diamonds) is compared with the corresponding cloak cell (circles with
  black outlines): \textbf{(a)}~the analytical transformation cloak and
  \textbf{(b)}~the optimised ($14\times10$) neural-field cloak.}
  \label{fig:bloch}
\end{figure}

As an independent diagnostic, we compute the Floquet–Bloch dispersion spectra of a unit cell comprising the entire triangular cloak cross-section shown in Fig.~\ref{fig:setup}(b). The cell is Bloch-periodic along the surface direction $x_1$,
traction-free at the top, and clamped at the bottom. 
Figure~\ref{fig:bloch} presents the resulting dispersion spectra for the uncloaked reference cell, the analytical transformation cloak, and the optimised ($14\times10$) neural-field cloak. Modes are classified using the inverse participation ratio (IPR)

\begin{equation}
\label{eq:ipr}
\mathrm{IPR}
=
|\psi^\star|\,
\frac{\sum_n A_n \lVert \mathbf{u}_n \rVert^4}
{\left(\sum_n A_n \lVert \mathbf{u}_n \rVert^2\right)^2},
\qquad
\lVert \mathbf{u}_n \rVert^2
=
|u_{1,n}|^2 + |u_{2,n}|^2,
\end{equation}
where $A_n$ is the nodal area and $|\psi^\star| = \sum_n A_n$. Modes with
$\mathrm{IPR}\geq2.5$ are identified as surface-localised and shown
as large outlined markers, whereas the remaining bulk modes are
shown as small dots. Marker colour represents the IPR. The analytical Rayleigh-wave branch, $f^\star=\xi$, folded at
each Brillouin-zone boundary, is included as a dashed reference
line.

For each configuration, $50$ Bloch wavenumbers and $400$ eigenmodes
were evaluated, giving $20{,}000$ modes per configuration. Within the range
$f^\star\leq2.5$, the numbers of surface-localised modes are $768$,
$725$, and $816$ for the reference, analytical-cloak, and
optimised-cloak cells, respectively.

\section{Microstructure-constrained realisation}\label{sec:micro}

The cloaks of Sec.~\ref{sec:multifreq} are tensor-valued: each cloak cell
carries a homogenised $(\mbf{p}_c,\rho)$ that need not correspond to
any manufacturable object. We now close the loop to an explicit, single-material
realisation, in which every cloak cell is filled with a two-phase solid/void
microstructure whose homogenised stiffness reproduces the optimised target, so
that the whole cloak can be built from one constituent by varying only the
internal pattern from cell to cell. The solid phase is concrete, taken as
isotropic with $E=30\,\text{GPa}$, $\nu=0.2$ and
$\rho=2300\,\text{kg}/\text{m}^3$ (plane strain), and the second phase is void;
the cloak is cast into the soil half-space of
Sec.~\ref{subsec:domain}. We first constrain the neural-field
optimisation to the manifold of realisable microstructures, then compare inverse design and conditional diffusion generative methods for turning
each per-cell target into an actual microstructure, and finally exhibit and
validate the realised cloak.

\subsection{Pipeline}\label{subsec:micro_pipeline}

\paragraph{Step 1 -- Microstructure dataset} A catalogue of $N\sim10^{6}$
binary $50\times50$ unit cells is generated by a squared-assembly cellular
automaton on a concrete/void substrate, following the cellular-automaton
generative approach of \citet{nakarmi2024}, who assemble a large,
topologically diverse library of mesostructures by evolving a pixel grid
under neighbour-count birth/death and survival rules.

\paragraph{Step 2 -- Homogenisation} 

Each unique cell is passed through standard
asymptotic homogenisation under periodic boundary
conditions~\citep{bendsoe1988,hassani1998,andreassen2014}, with the solid pixels
carrying the concrete moduli above and the void pixels an ersatz stiffness
$10^{-6}E$; the effective stiffness follows from the volume-averaged stress of
the unit-cell problems, the periodic boundary conditions guaranteeing
energetic consistency in the Hill--Mandel sense~\citep{hill1963}. The cell
problems are solved with the same JAX-FEM solver~\citep{xu2020jaxfem} as the
macro problem. Because the squared-assembly cells are
invariant under the point group $D_2$, the homogenised tensor has exactly the orthotropic
form~\eqref{eq:orthotropic} that is the design variable throughout the paper, so the descriptor carries the full (orthotropic) stiffness rather than an isotropic projection of it.

\paragraph{Step 3 -- Realisability prior}
A Gaussian mixture model (GMM) is fitted to the standardised
five-dimensional material descriptors of the catalogue. For each cell,
we define
\be\label{eq:standardised_descriptor}
  \mbf{z}
  =
  \operatorname{standardise}
  \left(
    \begin{bmatrix}
      \mbf{p}_c\\
      \rho
    \end{bmatrix}
  \right)
  \in\mathbb{R}^{5}.
\ee
Its probability density is
\be\label{eq:gmm_density}
  p_{\mathrm{GMM}}(\mbf{z})
  =
  \sum_{k=1}^{K}\pi_k\,
  \mcl{N}\!\left(\mbf{z}\mid\boldsymbol{\mu}_k,\boldsymbol{\Sigma}_k\right),
  \qquad
  \sum_{k=1}^{K}\pi_k=1,
\ee
where $K$ is the number of mixture components, $\pi_k$ their weights, and
$\boldsymbol{\mu}_k$ and $\boldsymbol{\Sigma}_k$ their means and covariance
matrices, respectively~\citep{mclachlan2000}. The realisable manifold is then
defined by a flat-top log-density threshold,
\be\label{eq:gmm_manifold}
  \mcl{M}
  =
  \left\{\mbf{z}:\log p_{\mathrm{GMM}}(\mbf{z})\ge\tau\right\},
  \qquad
  \tau
  =
  Q_q\!\left(
    \left\{\log p_{\mathrm{GMM}}(\mbf{z}_n)\right\}_{n=1}^{N}
  \right),
\ee
where $Q_q$ denotes the $q$-th empirical quantile over the catalogue.

\paragraph{Step 4 -- Constrained optimisation} The neural field of
Sec.~\ref{sec:method} is trained with the manifold penalty
\be\label{eq:gmm_pen}
  \mcl{L}_{\mathrm{GMM}}
  =
  \sum_i
  \max\!\bigl(
    0,\,
    \tau-\log p_{\mathrm{GMM}}(\mbf{z}_i)
  \bigr),
\ee
where $\mbf{z}_i$ is the standardised five-dimensional descriptor
constructed from $\mbf{p}_{c,i}$ and the density $\rho_i$ of cloak
cell $i$. The penalty is added to the single-frequency or band cloak
loss. The penalty pulls every cell
into $\mcl{M}$ without pinning it to a specific catalogue entry, while the
positive-definiteness and bounded-anisotropy guarantees of
Sec.~\ref{sec:constrained} keep every intermediate design admissible. For the
constrained runs we move from the $14\times10$ grid of Sec.~\ref{subsec:final14}
to a finer $20\times15$ grid  which places exactly $100$ cells inside the
cloak in the expectation that the extra per-cell freedom would let the
constrained optimisation reach a good realisable solution more easily.

\paragraph{Step 5 -- Microstructure realisation} Each optimised per-cell target $(\mbf{p}_c,\rho)$ is turned into an explicit $50\times50$ solid/void cell by one of the two generative methods of Sec.~\ref{subsec:micro_methods} (or a nearest-neighbour dataset baseline).

\paragraph{Step 6 -- Evaluation} The macro FEM is re-solved with every cloak
cell carrying the homogenised stiffness and density of its realised microstructure, and the
cloak ratio $\eta$~\eqref{eq:cloak_ratio} is recomputed.

\subsection{Two realisation methods}\label{subsec:micro_methods}

Both methods solve the same per-cell inverse problem -- given a target
orthotropic stiffness and density, produce a binary $50\times50$ cell whose
homogenisation reproduces it -- but approach it from opposite directions.

\paragraph{Inverse design (neural field)} A coordinate network parameterises a
continuous density field over the $50\times50$ cell. It is relaxed during
training and driven to a binary pattern through a soft-to-hard schedule, with a
connectivity penalty that suppresses disconnected concrete islands and floating
voids. The cell's periodic-FEM homogenised stiffness is differentiable in the
network weights, so the pattern is optimised directly against the per-cell target
error. The method is deterministic and returns a single, locally optimal cell for
each target.

\paragraph{Diffusion (conditional generative)} A conditional denoising diffusion
model is trained on the catalogue to sample binary cells conditioned on
the target stiffness parameters $\mbf{p}_c$ and density $\rho$, following the guided-diffusion inverse-design approach of \citet{yang2024}. Because sampling is stochastic, $N$ candidates are drawn per cell and the one whose homogenised stiffness is closest to the target is kept (\emph{best-of-$N$}). The generative prior keeps every sample
inside the manifold of manufacturable cells while exploring a diverse set of
realisations for each target.

\begin{figure}[htbp]
  \centering
  \includegraphics[width=\linewidth]{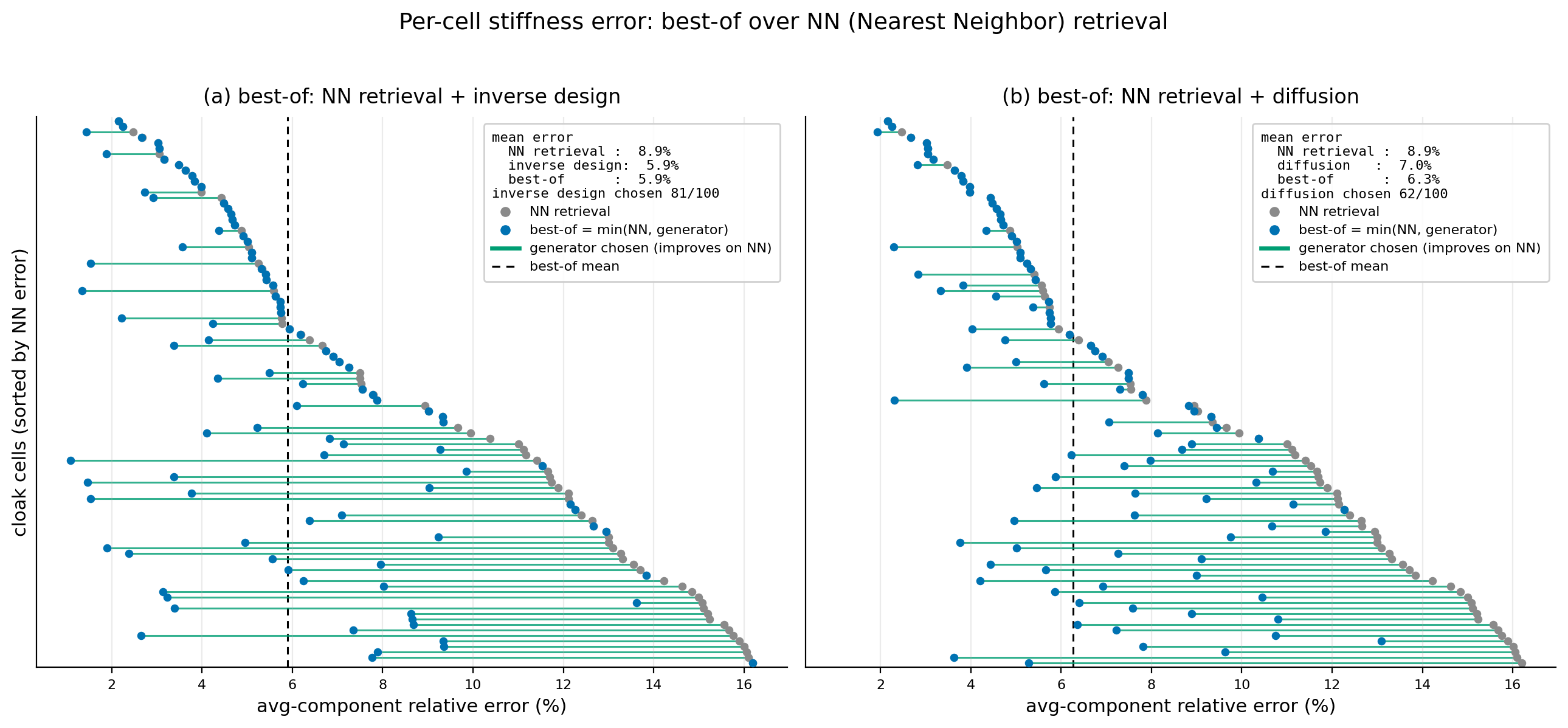}
  \caption{Per-cell stiffness-matching error over the $100$ cloak cells for the
  two realisation methods against NN (Nearest Neighbor) retrieval: (a)~NN +
  inverse design, (b)~NN + diffusion. Each row is a cell; the bar links the NN
  error (grey) to the per-cell best-of selection (blue, the lower-error of the
  two). Best-of lowers the mean avg-component stiffness error from $8.9\%$ to
  $5.9\%$ (inverse) and $6.3\%$ (diffusion); cells are sorted by NN error.}
  \label{fig:micro_bestof_dumbbell}
\end{figure}

Figure~\ref{fig:micro_bestof_dumbbell} quantifies how well each method improves its
per-cell target across the $100$ cloak cells compared to the nearest-neighbour snap. 
Section~\ref{subsec:micro_results} shows, however, this per-cell stiffness accuracy 
does not translate directly into whole-cloak performance.

\subsection{Results}\label{subsec:micro_results}

Table~\ref{tab:micro_methods} reports the cloak ratio at $f^\star=2.0$ on the
$20\times15$ macro grid ($100$ cloak cells) for the three realisation
strategies, each evaluated with the homogenised stiffness of the realised cells. The continuous orthotropic Cauchy optimum reaches $\eta=0.99$. When
the generated microstructures are represented by their homogenised
stiffness and density, the nearest-neighbour dataset snap retains
$\eta=0.82$, the neural-field inverse design reaches $\eta=0.94$, and
the best-of-$N$ diffusion realisation reaches $\eta=0.967$, within
$3\%$ of the continuous optimum.
\begin{table}[htbp]
\centering
\caption{Microstructure realisation of the constrained $20\times15$ cloak at
$f^\star=2.0$ ($100$ cloak cells): cloak ratio $\eta$ (homogenised, matched) for
the three cell-realisation strategies, against the continuous orthotropic Cauchy optimum and the
ideal polar cloak. Best-of-$N$ diffusion recovers $\eta=0.97$.}
\label{tab:micro_methods}
\begin{tabular*}{\textwidth}{@{\extracolsep{\fill}}lc@{}}
\toprule
Realisation of the $100$ cloak cells & $\eta$ \\
\midrule
Continuous orthotropic Cauchy optimum (homogenised, GMM-constrained)          & $0.990$ \\
\midrule
Nearest-neighbour dataset match                                & $0.816$ \\
Inverse design (neural field)                                  & $0.936$ \\
\textbf{Diffusion (best-of-$N$)}                               & $\mathbf{0.967}$ \\
\midrule
Ideal continuous polar $\mbf{c}_{\mathrm{eff}}$ & $0.999$ \\
\bottomrule
\end{tabular*}
\end{table}

\begin{figure}[htbp]
  \centering
  \includegraphics[width=0.9\linewidth]{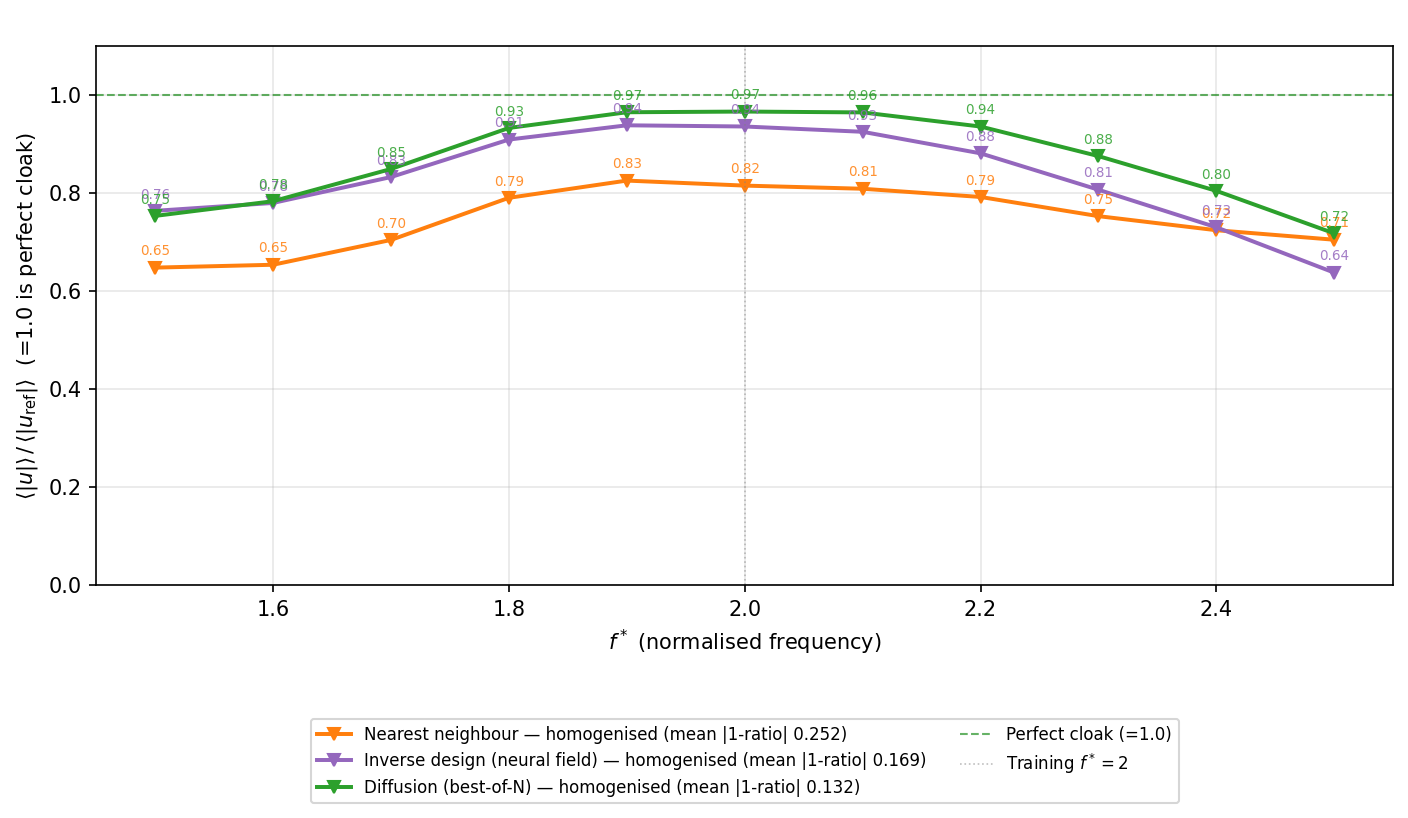}
  \caption{Frequency sweep of the cloak ratio
  $\eta(f)=\langle|\mbf{u}|\rangle/\langle|\mbf{u}_{\mathrm{ref}}|\rangle$ for the
  three microstructure-realisation methods on the matched homogenised
  $20\times15$ cloak: nearest-neighbour dataset snap (orange), neural-field
  inverse design (purple), and best-of-$N$ diffusion (green). Diffusion stays
  closest to the perfect-cloak line ($\eta=1$) across the band and has the
  smallest mean distortion; the dashed line is the training frequency
  $f^\star=2$.}
  \label{fig:micro_method_sweep}
\end{figure}

The advantage of the generative realisation is not confined to the design
frequency. Figure~\ref{fig:micro_method_sweep} sweeps $\eta(f)$ across
$f^\star\in[1.5,2.5]$ for the three methods on the matched homogenised cloak:
diffusion holds $\eta\gtrsim0.93$ over $f^\star\in[1.8,2.2]$ and has the smallest
mean in-band distortion ($\langle|1-\eta|\rangle=0.13$, against $0.17$ for
inverse design and $0.25$ for nearest neighbour), so better per-cell realisation
improves the cloak across the whole band, not only at $f^\star=2$. We note that a
greedy per-cell hybrid  picking, for each cell independently, whichever of the
two methods best matches that cell's target does \emph{not} beat diffusion
($\eta=0.95$): minimising per-cell stiffness error is not the same as maximising
whole-cloak cloaking, because the ratio couples the cells through the shared wave
field.

\subsection{Realised cloak and validation}\label{subsec:micro_realisation}

\begin{figure}[htbp]
  \centering
  \begin{subfigure}{0.49\textwidth}
    \centering
    \includegraphics[width=\linewidth]{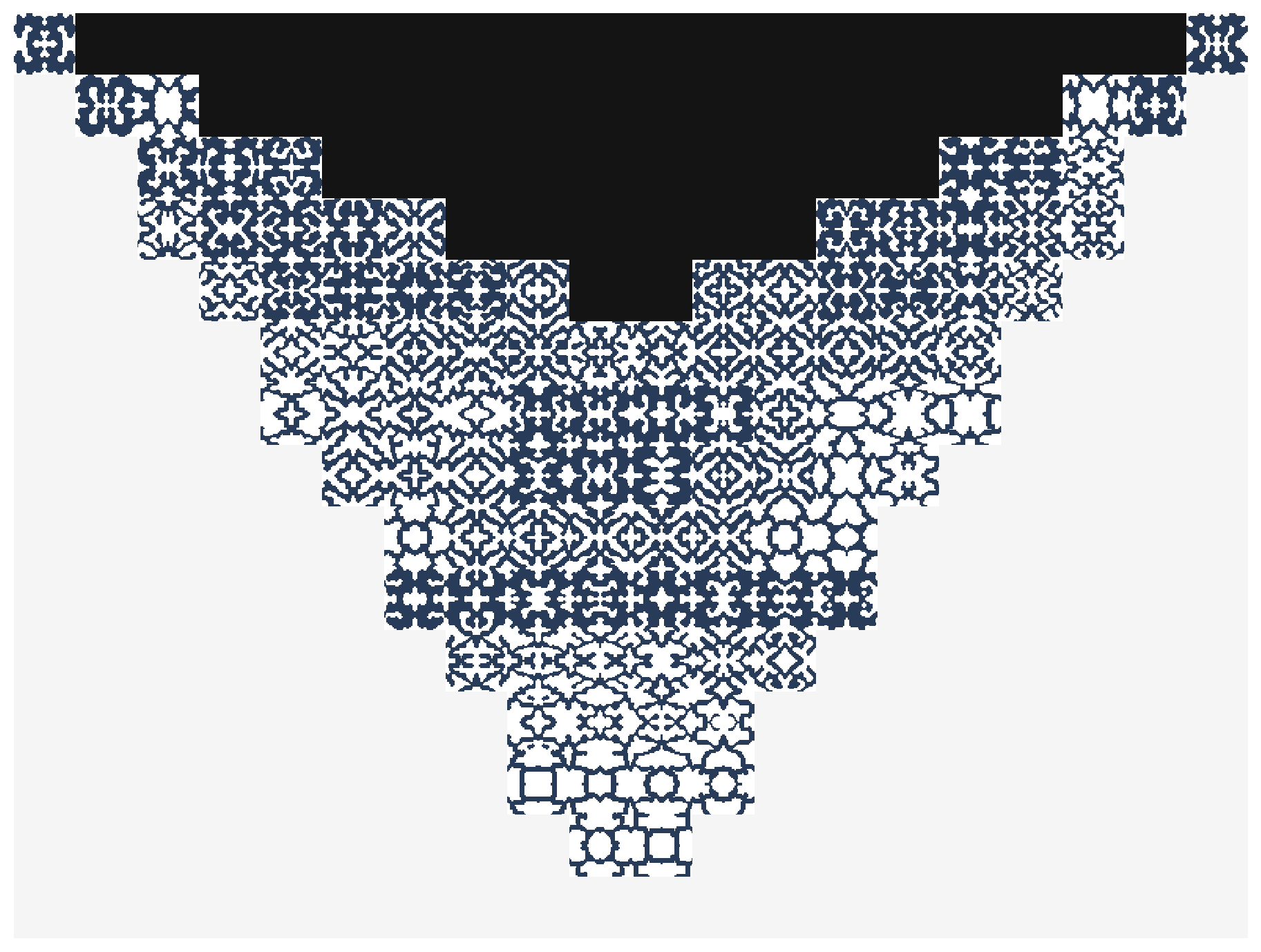}
    \caption{Diffusion (best-of-$N$)}
    \label{fig:micro_tiling_diff}
  \end{subfigure}\hfill
  \begin{subfigure}{0.49\textwidth}
    \centering
    \includegraphics[width=\linewidth]{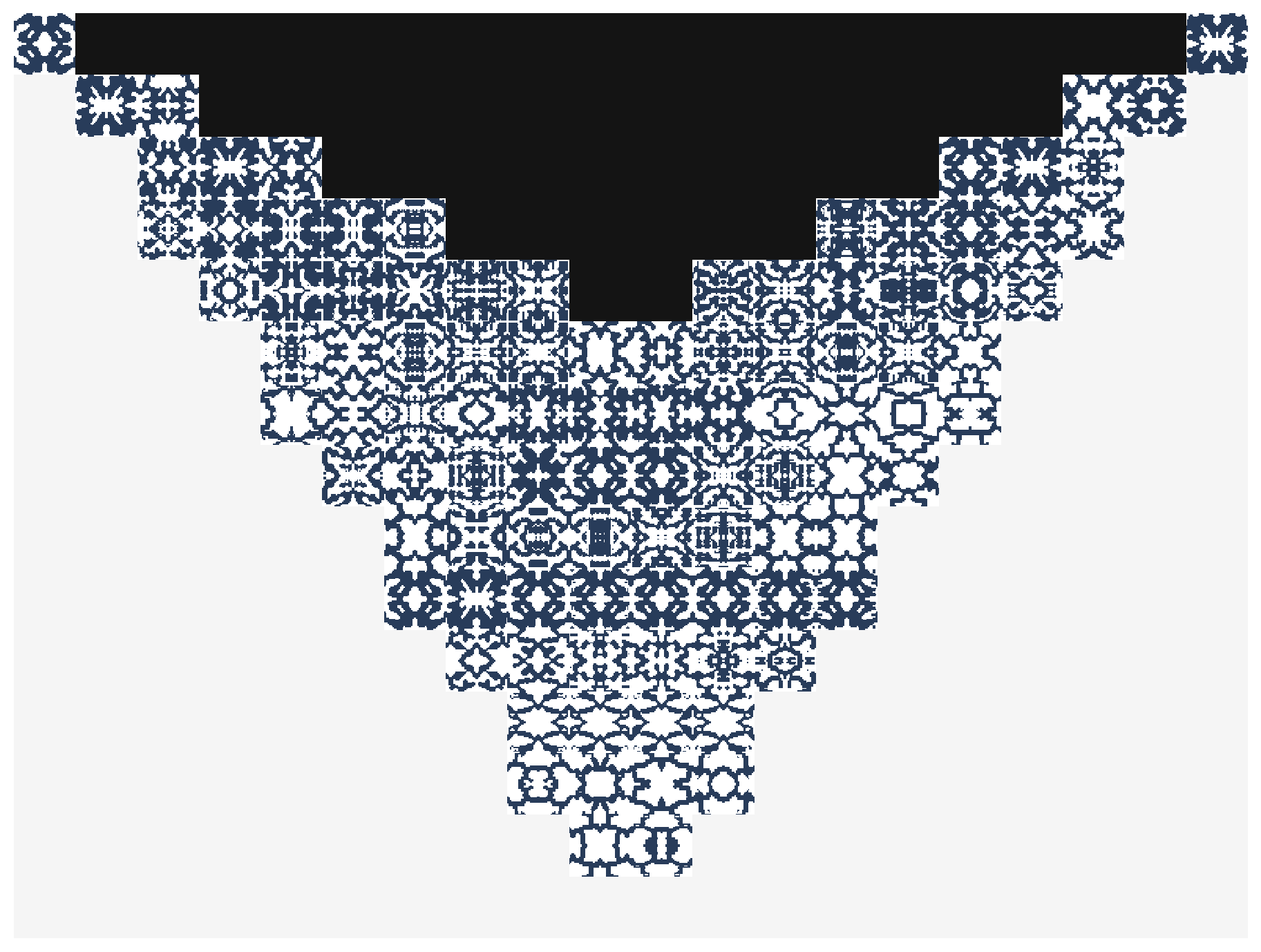}
    \caption{Inverse design (neural field)}
    \label{fig:micro_tiling_inv}
  \end{subfigure}
  \caption{Explicit realisation of the triangular carpet cloak from a single
  concrete phase: the $100$ generated $50\times50$ unit cells tiled at the
  macro-cell positions they occupy in the cloak. Concrete is dark, void white; the
  black inverted triangle is the surface notch and the light grey region is the
  surrounding soil half-space. (a)~Best-of-$N$ diffusion and (b)~neural-field inverse
  design produce visibly different but equally manufacturable patterns for the
  same optimised per-cell targets.}
  \label{fig:micro_tiling}
\end{figure}

Figure~\ref{fig:micro_tiling} shows the explicit solid/void realisation of the
cloak: the $100$ generated cells tiled at the macro-cell positions
they occupy inside the triangular carpet, with the surface notch and the
surrounding soil half-space. The whole cloak is a single concrete phase perforated by a
cell-varying void pattern, densest near the outer boundary and most strongly
perforated near the sheared defect tip, mirroring the depth variation of the
ideal tensor~\eqref{eq:Ceff}.

Two validation levels are considered. In the homogenised validation,
each generated unit cell is replaced by its effective orthotropic
stiffness and density, giving a cloak ratio of $\eta=0.967$ for the
best-of-$N$ diffusion cloak. In the fully resolved validation, the
generated solid/void geometries themselves are inserted into the cloak
region and meshed explicitly. The concrete phase is discretised with
quadratic triangular elements, while the void boundaries are treated as
traction-free surfaces. At the design frequency $f^\star=2.0$, the
fully resolved cloak yields a cloak ratio of approximately
$\eta=0.76$.

\section{Discussion}\label{sec:discussion}
The three regimes above share a single architectural choice, a coordinate conditioned neural field on top of a differentiable FEM, and they show, in different forms, the same advantage over handcrafted symmetrisation strategies. In the single frequency case the network finds a low-distortion realisable cloak by directly searching four orthotropic Cauchy moduli per cell; the broken minor symmetry belongs to the ideal polar tensor and represents the gap that any realisable Cauchy design (ours included) must approximate. In
the multifrequency case the network's implicit smoothness eliminates
the spurious cell-to-cell oscillations that plague raw per-cell
optimisation and the band objective flattens the loss across a wide
target band. In the microstructure-constrained case the same network
output is projected onto a realisability manifold and realised as an
explicit single-material microstructure. When the generated cells are represented by their homogenised properties, the best-of-$N$ diffusion cloak reaches $\eta=0.967$, showing that the microstructures reproduce the continuum targets with little loss at the effective-property level. Direct simulation of the fully resolved solid/void cloak gives $\eta\approx0.76$. Thus, matching the quasi-static homogenised elastic constants is not sufficient to reproduce the complete finite-frequency response of the assembled cloak. Nevertheless, the explicit model retains a substantial cloaking effect, demonstrating that the continuum-to-microstructure design survives full geometrical realisation.

\paragraph{Future work} The same differentiable pipeline, coupling a
neural field to a finite-element solver, applies to other Rayleigh-wave
design problems beyond carpet cloaking, such as wave focusing,
waveguiding, and vibration isolation, by changing only the objective. On
the realisability side, we have already replaced the discrete database
with a differentiable generative model of the realisable manifold,
queried during optimisation; this model still generates one unit cell at a
time, and producing a full multiscale image of the cloak, where
neighbouring cells connect and vary smoothly across the domain, is an
open problem that may again be a good fit for diffusion models. Two further extensions stand out. The first is to move from the
two-dimensional carpet cloak to a three-dimensional cloak. The
differentiable JAX-FEM pipeline carries over, although the cost of the
full solve grows sharply. The second is to move beyond the orthotropic
Cauchy class used here. The ideal polar tensor breaks minor symmetry,
so its realisation requires Cosserat or polar microstructures and a
homogenisation framework capable of representing the corresponding
generalised elastic response.

\section{Conclusion}\label{sec:conclusion}

We have presented a unified, differentiable design optimisation pipeline for
2D Rayleigh-wave carpet cloaks in which the per-cell
material is parameterised by orthotropic stiffness parameters $\mbf{p}_c$ and a scalar density $\rho$, searched by a coordinate-conditioned neural field through an end-to-end differentiable JAX-FEM pipeline. This is an alternative to closed-form arithmetic-mean
symmetrisation of the ideal polar tensor that directly explores the
realisable Cauchy design space. The pipeline achieves near-perfect cloaking and flattens the cloak loss across the target frequency band. In a subsequent step, we have compared diffusion and neural-field optimisation for projecting these continuum targets onto a realisability manifold, with diffusion performing better, and when the selected microstructures are represented through their homogenised effective stiffness and density, the resulting cloak retains a high performance of $\eta=0.97$. Most importantly, direct FEM simulation of the fully resolved microstructured cloak still achieves approximately $\eta=0.76$, demonstrating that a substantial cloaking effect survives the transition from effective continuum properties to an explicit single-material architecture. This provides a practical route towards physically realisable Rayleigh-wave cloaks and further multiscale optimisation of their dynamic performance.

\bibliographystyle{elsarticle-num-names}
\bibliography{cas-refs}

@article{chatzopoulos2023,
  author  = {Chatzopoulos, Z. and Palermo, A. and Diatta, A. and Guenneau, S. and Marzani, A.},
  title   = {Cloaking {Rayleigh} waves via symmetrized elastic tensors},
  journal = {Int. J. Eng. Sci.},
  volume  = {191},
  pages   = {103899},
  year    = {2023},
  doi     = {10.1016/j.ijengsci.2023.103899}
}

@article{nassar2018,
  author  = {Nassar, H. and Chen, Y. Y. and Huang, G. L.},
  title   = {A degenerate polar lattice for cloaking in full two-dimensional elastodynamics and statics},
  journal = {Proc. R. Soc. A},
  volume  = {474},
  pages   = {20180523},
  year    = {2018},
  doi     = {10.1098/rspa.2018.0523}
}

@article{buckmann2014elasto,
  title={An elasto-mechanical unfeelability cloak made of pentamode metamaterials},
  author={B{\"u}ckmann, Tiemo and Thiel, Michael and Kadic, Muamer and Schittny, Robert and Wegener, Martin},
  journal={Nature communications},
  volume={5},
  number={1},
  pages={4130},
  year={2014},
  publisher={Nature Publishing Group UK London}
}

@article{diatta2014controlling,
  title={Controlling solid elastic waves with spherical cloaks},
  author={Diatta, Andre and Guenneau, Sebastien},
  journal={Applied Physics Letters},
  volume={105},
  number={2},
  year={2014},
  publisher={AIP Publishing}
}

@article{kadic2020elastodynamic,
  title={Elastodynamic behavior of mechanical cloaks designed by direct lattice transformations},
  author={Kadic, Muamer and Wegener, Martin and Nicolet, Andr{\'e} and Zolla, Fr{\'e}d{\'e}ric and Guenneau, S{\'e}bastien and Diatta, Andr{\'e}},
  journal={Wave Motion},
  volume={92},
  pages={102419},
  year={2020},
  publisher={Elsevier}
}

@article{quadrelli2021elastic,
  title={Elastic wave near-cloaking},
  author={Quadrelli, Davide Enrico and Craster, Richard and Kadic, Muamer and Braghin, Francesco},
  journal={Extreme Mechanics Letters},
  volume={44},
  pages={101262},
  year={2021},
  publisher={Elsevier}
}

@article{brun2009,
  author  = {Brun, M. and Guenneau, S. and Movchan, A. B.},
  title   = {Achieving control of in-plane elastic waves},
  journal = {Appl. Phys. Lett.},
  volume  = {94},
  pages   = {061903},
  year    = {2009},
  doi     = {10.1063/1.3068491}
}

@article{norris2011,
  author  = {Norris, A. N. and Shuvalov, A. L.},
  title   = {Elastic cloaking theory},
  journal = {Wave Motion},
  volume  = {48},
  pages   = {525--538},
  year    = {2011},
  doi     = {10.1016/j.wavemoti.2011.03.002}
}

@article{wang2022,
  author  = {Wang, Liwei and Boddapati, Jagannadh and Liu, Ke and Zhu, Ping and Daraio, Chiara and Chen, Wei},
  title   = {Mechanical cloak via data-driven aperiodic metamaterial design},
  journal = {Proc. Natl. Acad. Sci. U.S.A.},
  volume  = {119},
  number  = {13},
  pages   = {e2122185119},
  year    = {2022},
  doi     = {10.1073/pnas.2122185119}
}

@article{milton1995,
  author  = {Milton, G. W. and Cherkaev, A. V.},
  title   = {Which elasticity tensors are realizable?},
  journal = {J. Eng. Mater. Technol.},
  volume  = {117},
  pages   = {483--493},
  year    = {1995},
  doi     = {10.1115/1.2804743}
}

@article{milton2006,
  author  = {Milton, G. W. and Briane, M. and Willis, J. R.},
  title   = {On cloaking for elasticity and physical equations with a transformation invariant form},
  journal = {New J. Phys.},
  volume  = {8},
  pages   = {248},
  year    = {2006},
  doi     = {10.1088/1367-2630/8/10/248}
}

@article{pendry2006,
  author  = {Pendry, J. B. and Schurig, D. and Smith, D. R.},
  title   = {Controlling Electromagnetic Fields},
  journal = {Science},
  volume  = {312},
  pages   = {1780--1782},
  year    = {2006},
  doi     = {10.1126/science.1125907}
}

@article{li2008,
  author  = {Li, J. and Pendry, J. B.},
  title   = {Hiding under the carpet: a new strategy for cloaking},
  journal = {Phys. Rev. Lett.},
  volume  = {101},
  pages   = {203901},
  year    = {2008},
  doi     = {10.1103/PhysRevLett.101.203901}
}

@article{buckmann2015,
  author  = {B{\"u}ckmann, T. and Kadic, M. and Schittny, R. and Wegener, M.},
  title   = {Mechanical cloak design by direct lattice transformation},
  journal = {Proc. Natl. Acad. Sci. U.S.A.},
  volume  = {112},
  number  = {16},
  pages   = {4930--4934},
  year    = {2015},
  doi     = {10.1073/pnas.1501240112}
}

@article{willis1981,
  author  = {Willis, J. R.},
  title   = {Variational principles for dynamic problems for inhomogeneous elastic media},
  journal = {Wave Motion},
  volume  = {3},
  pages   = {1--11},
  year    = {1981},
  doi     = {10.1016/0165-2125(81)90008-1}
}

@article{norris2008,
  author  = {Norris, A. N.},
  title   = {Acoustic cloaking theory},
  journal = {Proc. R. Soc. A},
  volume  = {464},
  pages   = {2411--2434},
  year    = {2008},
  doi     = {10.1098/rspa.2008.0076}
}

@article{nakarmi2024,
  author  = {Nakarmi, Sushan and Leiding, Jeffery A. and Lee, Kwan-Soo and Daphalapurkar, Nitin P.},
  title   = {Predicting non-linear stress--strain response of mesostructured cellular materials using supervised autoencoder},
  journal = {Computer Methods in Applied Mechanics and Engineering},
  year    = {2024},
  volume  = {432},
  pages   = {117372},
  doi     = {10.1016/j.cma.2024.117372}
}

@article{yang2024,
  author  = {Yang, Yanyan and Wang, Lili and Zhai, Xiaoya and Chen, Kai and Liu, Weiming and Fang, Ming and Xu, Tuanjie and Belyaev, Alexander and Wang, Charlie C. L. and Liu, Ligang},
  title   = {Guided Diffusion for Fast Inverse Design of Density-based Mechanical Metamaterials},
  journal = {arXiv preprint arXiv:2401.13570},
  year    = {2024}
}

@inproceedings{sitzmann2020siren,
  author    = {Sitzmann, Vincent and Martel, Julien N. P. and Bergman, Alexander W. and Lindell, David B. and Wetzstein, Gordon},
  title     = {Implicit Neural Representations with Periodic Activation Functions},
  booktitle = {Advances in Neural Information Processing Systems (NeurIPS)},
  year      = {2020}
}

@article{mildenhall2020nerf,
  author  = {Mildenhall, Ben and Srinivasan, Pratul P. and Tancik, Matthew and Barron, Jonathan T. and Ramamoorthi, Ravi and Ng, Ren},
  title   = {{NeRF}: Representing Scenes as Neural Radiance Fields for View Synthesis},
  journal = {Communications of the ACM},
  volume  = {65},
  number  = {1},
  pages   = {99--106},
  year    = {2021},
  doi     = {10.1145/3503250}
}

@article{tancik2020fourier,
  author  = {Tancik, Matthew and Srinivasan, Pratul P. and Mildenhall, Ben and Fridovich-Keil, Sara and Raghavan, Nithin and Singhal, Utkarsh and Ramamoorthi, Ravi and Barron, Jonathan T. and Ng, Ren},
  title   = {Fourier features let networks learn high frequency functions in low dimensional domains},
  journal = {Advances in Neural Information Processing Systems},
  volume  = {33},
  pages   = {7537--7547},
  year    = {2020}
}

@article{hoyer2019,
  author  = {Hoyer, Stephan and Sohl-Dickstein, Jascha and Greydanus, Sam},
  title   = {Neural reparameterization improves structural optimization},
  journal = {arXiv preprint arXiv:1909.04240},
  year    = {2019}
}

@article{chandrasekhar2021,
  author  = {Chandrasekhar, Aaditya and Suresh, Krishnan},
  title   = {{TOuNN}: Topology Optimization using Neural Networks},
  journal = {Struct. Multidiscip. Optim.},
  volume  = {63},
  number  = {3},
  pages   = {1135--1149},
  year    = {2021},
  doi     = {10.1007/s00158-020-02748-4}
}

@inproceedings{zehnder2021,
  author    = {Zehnder, Jonas and Li, Yue and Coros, Stelian and Thomaszewski, Bernhard},
  title     = {{NTopo}: Mesh-free Topology Optimization using Implicit Neural Representations},
  booktitle = {Advances in Neural Information Processing Systems (NeurIPS)},
  volume    = {34},
  year      = {2021}
}

@article{xu2020jaxfem,
  author  = {Xue, Tianju and Liao, Shuheng and Gan, Zhengtao and Park, Chanwook and Xie, Xiaoyu and Liu, Wing Kam and Cao, Jian},
  title   = {{JAX-FEM}: A differentiable {GPU}-accelerated 3{D} finite element solver for automatic inverse design and mechanistic data science},
  journal = {Comput. Phys. Commun.},
  volume  = {291},
  pages   = {108802},
  year    = {2023},
  doi     = {10.1016/j.cpc.2023.108802}
}

@article{bendsoe1988,
  author  = {Bends{\o}e, M. P. and Kikuchi, N.},
  title   = {Generating optimal topologies in structural design using a homogenization method},
  journal = {Comput. Methods Appl. Mech. Engrg.},
  volume  = {71},
  pages   = {197--224},
  year    = {1988},
  doi     = {10.1016/0045-7825(88)90086-2}
}

@article{hill1963,
  author  = {Hill, R.},
  title   = {Elastic properties of reinforced solids: Some theoretical principles},
  journal = {J. Mech. Phys. Solids},
  volume  = {11},
  pages   = {357--372},
  year    = {1963},
  doi     = {10.1016/0022-5096(63)90036-X}
}

@article{hassani1998,
  author  = {Hassani, B. and Hinton, E.},
  title   = {A review of homogenization and topology optimization {I} -- homogenization theory for media with periodic structure},
  journal = {Comput. Struct.},
  volume  = {69},
  pages   = {707--717},
  year    = {1998},
  doi     = {10.1016/S0045-7949(98)00131-X}
}

@article{andreassen2014,
  author  = {Andreassen, Erik and Andreasen, Casper Schousboe},
  title   = {How to determine composite material properties using numerical homogenization},
  journal = {Comput. Mater. Sci.},
  volume  = {83},
  pages   = {488--495},
  year    = {2014},
  doi     = {10.1016/j.commatsci.2013.09.006}
}

@book{mclachlan2000,
  author    = {McLachlan, Geoffrey J. and Peel, David},
  title     = {Finite Mixture Models},
  publisher = {John Wiley \& Sons},
  year      = {2000},
  doi       = {10.1002/0471721182}
}

\end{document}